\documentclass[usenatbib]{mn2e}
\usepackage{amsmath}
\usepackage{natbib}
\usepackage{epsfig}
\usepackage{txfonts}
\usepackage{pict2e}
\usepackage{color}

\newcommand{\zmuy}{{z_{\mu,y}}}

\newcommand{\Mpc}{{\rm Mpc}}

\newcommand{\expf}[1]{{{\rm e}^{#1}}}

\newcommand{\nS}{n_{\rm S}}

\newcommand{\vgh}{{\hat{\boldsymbol\gamma}}}
\newcommand{\vghp}{{\hat{\boldsymbol\gamma}'}}

\newcommand{\taudot}{\dot{\tau}}
\newcommand{\kD}{k_{\rm D}}
\newcommand{\rs}{r_{\rm s}}

\newcommand{\id}{{\,\rm d}}

\newcommand{\beq}{\begin{equation}}   %

\newcommand{\eeq}{\end{equation}}   %

\newcommand{\beqa}{\begin{eqnarray}}   %

\newcommand{\eeqa}{\end{eqnarray}}   %

\newcommand{\beal}{\begin{align}}

\newcommand{\enal}{\end{align}}

\newcommand{\bspl}{\begin{split}}

\newcommand{\espl}{\end{split}}

\newcommand{\bsub}{\begin{subequations}}

\newcommand{\esub}{\end{subequations}}

\newcommand{\bmulti}{\begin{multline}}   %

\newcommand{\beqm}{\begin{mathletters}}   %

\newcommand{\eeqm}{\end{mathletters}}   %

\newcommand{\Ne}{N_{\rm e}}

\newcommand{\Tg}{T_{\gamma}}

\newcommand{\sigT}{\sigma_{\rm T}}

\newcommand{\vek} [1]{{\mathbfit #1}}

\newcommand{\pot}[2]{#1 \times 10^{#2}}

\newcommand{\Trans}{{\mathcal{T}}}

\definecolor{redwine}{rgb}{.72,0,0}

\usepackage{hyperref}
\usepackage{grffile}
\usepackage{graphics}
\usepackage{booktabs}

\title[CMB distortion anisotropies]
{Evolution of CMB spectral distortion anisotropies and tests of primordial non-Gaussianity}

\author[Chluba et al.]{
Jens~Chluba$^{1}$\thanks{E-mail:Jens.Chluba@Manchester.ac.uk}, Emanuela Dimastrogiovanni$^{2}$, 
Mustafa A. Amin$^{3}$, Marc Kamionkowski$^{4}$
\\
$^{1}$ Jodrell Bank Centre for Astrophysics, University of Manchester, Oxford Road, Manchester M13 9PL, UK
\\
$^{2}$ Department of Physics and School of Earth and Space Exploration Arizona State University, Tempe, AZ 85827, USA
\\
$^{3}$ Physics \& Astronomy Department, Rice University, Houston, Texas 77005-1827, USA
\\
$^{4}$ Department of Physics and Astronomy, Johns Hopkins University, 3400 N. Charles St., Baltimore, MD 21218, USA
}

\begin{document}

\date{\vspace{-6mm}{Accepted 2016 December 8. Received 2016 October 27}}

\maketitle

\begin{abstract}
Anisotropies in distortions to the frequency spectrum of the cosmic microwave background (CMB) can be created through spatially varying heating processes in the early Universe. For instance, the dissipation of small-scale acoustic modes does create distortion anisotropies, in particular for non-Gaussian primordial perturbations. In this work, we derive approximations that allow describing the associated distortion field. We provide a systematic formulation of the problem using Fourier-space window functions, clarifying and generalizing previous approximations. Our expressions highlight the fact that the amplitudes of the spectral-distortion fluctuations induced by non-Gaussianity depend also on the homogeneous value of those distortions. Absolute measurements are thus required to obtain model-independent distortion constraints on primordial non-Gaussianity. We also include a simple description for the evolution of distortions through photon diffusion, showing that these corrections can usually be neglected. Our formulation provides a systematic framework for computing higher order correlation functions of distortions with CMB temperature anisotropies and can be extended to describe correlations with polarization anisotropies.
\end{abstract}


\section{Introduction}
\label{sec:Intro}
Measurements of cosmic microwave background (CMB) spectral distortions can teach us about the thermal history of the Universe \citep[see][for recent overviews]{Chluba2011therm, Sunyaev2013, Tashiro2014, Chluba2016}. Aside from the distortion signals imprinted at arcminute angular scales by clusters of galaxies through the Sunyaev-Zeldovich (SZ) effect \citep{Zeldovich1969, Carlstrom2002}, anisotropies in the distortion to the CMB frequency spectrum are expected to be small. However, anisotropic heating caused by {\it non-standard} early-universe processes can in principle lead to observable distortion anisotropies \citep{Chluba2012}. One example is related to the dissipation of small-scale perturbations (wavenumber $k\simeq 10^2-10^4\,\Mpc^{-1}$) in the photon-baryon fluid in the presence of ultra-squeezed limit non-Gaussianity \citep{Pajer2012, Ganc2012}. In this case, the local acoustic heating rate is modulated by large-scale modes ($k\simeq 10^{-3}-10^{-2}\,\Mpc^{-1}$), such that the CMB spectrum can vary across that sky. This effect can be used to place limits on primordial non-Gaussianity and test its scale-dependence \citep{Biagetti2013, Emami2015, Ema2016} using future spectrometers like {\it PIXIE} \citep{Kogut2011, Kogut2016SPIE}. 

Several estimates for the observability of these signals through their correlation with large-scale temperature perturbations can be found in the literature \citep{Pajer2012, Ganc2012, Biagetti2013, Emami2015, Khatri2015, Ota2015muiso, Creque2016}. In addition, new estimates related to damping-induced acoustic reheating \citep{Naruko2015}, correlations with polarization anisotropies \citep{Ota2016} and higher order correlation functions \citep{Bartolo2016, Shiraishi2016} as a new probe of primordial non-Gaussianity have been studied. All the aforementioned computations are based on different approximations and assumptions. Here, we develop a common formulation of the problem using Fourier-space window functions. These functions allow separating dissipation and thermalization physics from the statistical properties of the primordial perturbations and their spatial evolution in a transparent way with minimal assumptions. 

We use our formulation to justify the simple approximations given recently in \citet{Emami2015} for the primordial contributions to the chemical potential ($\mu$) and Compton-$y$ correlation functions with the CMB temperature, $C^{\mu T}_\ell$ and $C^{yT}_\ell$. These approximations directly show that the cross power spectra not only depend on the level of non-Gaussianity at small scales ($k\simeq 10^3\,\Mpc^{-1}$), parametrized by $f_{\rm NL}$, but also on the amplitude of the average distortion signal created by the homogenous heating term. This shows that there are two independent ways to enhance the amplitude of the primordial distortion fluctuations: by larger non-Gaussianity and/or a modified small-scale power spectrum. This highlights that independent limits on $f_{\rm NL}$ can only be derived in absolute measurements (e.g., {\it PIXIE}), for which the average distortion of the CMB monopole is also obtained, while in differential measurements (e.g., {\it CORE, Litebird}) the interpretation remains model-dependent. 

We explicitly model the transition between $\mu$ and $y$-type distortions \citep{Sunyaev1970mu, Zeldovich1969} using {\it distortion visibility functions} \citep{Chluba2013Green, Chluba2016}. We also include the effect of photon diffusion to the transfer functions for the distortion anisotropies. We highlight that, in contrast to temperature perturbations, no pressure waves appear for distortions. Distortion anisotropies simply smear out before propagating wave fronts can develop, with a damping scale that is $\simeq \sqrt{15/8}\simeq 1.4$ times smaller than the Silk-damping scale for temperature perturbations. This modification is usually not included and only becomes noticeable at multipoles $\ell \gtrsim 200$.

We compare the results for different approximations and also improve previous analytic expressions capturing the $\ell$-dependence of the $\mu-T$ correlation function caused by temperature transfer effects more accurately. We provide simple analytic expressions for the distortion signals, which can be easily evaluated by specifying the small-scale power spectrum and scale-dependence of $f_{\rm NL}$. Overall the main goal is to clarify and simplify the formulation of the problem for anisotropic heating processes. The expressions can be extended to the case of polarization and higher order correlation functions in a straightforward manner, but this is left to future work.

The paper is structured as follows: in Sect.~\ref{anisotropies}, we give the formulation of the problem and compute the temperature-distortion correlation functions. We directly compare with previous numerical estimates in Sect.~\ref{sec:results_muT}. In Sect.~\ref{transfer_effects}, we discuss distortion transfer effects, showing that they are small at $\ell \lesssim 200$. Our conclusions are presented in Sect.~\ref{sec:conclusions}.

\vspace{-3mm}
\section{Anisotropic spectral distortions through damping of small-scale perturbations} 
\label{anisotropies}
Following the evolution of the CMB spectrum with anisotropies in general is a hard problem. Not only spatial photon diffusion, but also redistribution of photons in energy (Compton scattering) and photon production (Bremsstrahlung and double Compton emission) have to be included for an anisotropic medium. A formulation of the required evolution equations was given by \citet{Chluba2012}. Here, we shall start by neglecting spatial transfer effects for the distortion evolution before decoupling at $z_{\rm rec}\approx 10^3$. We will return to the more general problem in Sect.~\ref{transfer_effects}. Under this assumption, we only need to specify the spatially varying heating rate caused by the damping of primordial temperature perturbations, which is given by the angular average 
\begin{equation}\label{eq0}
\frac{\id}{\id t}\left[\frac{Q(\vek{x},t)}{\rho_{\gamma}}\right]
\approx-4\langle \Theta \dot{\Theta}\rangle_{\Omega}
\equiv
-4\int \Theta\dot{\Theta}\,\frac{\id^{2}\hat{n}}{4\pi}
\end{equation}
over different directions $\hat{\vek{n}}$. Here, $\Theta$ denotes the CMB temperature fluctuation at any location $\vek{x}$. We only compute the monopole of the heating rate (angle-average), but do not consider the dipolar and quadrupolar heating rate, which cause tiny corrections\footnote{This can be seen by considering the spherical harmonic coefficients of the effective heating rate, $a_{\ell m}^{\dot Q}=-4\int \Theta\dot{\Theta}\,Y^*_{\ell m}(\hat{\vek{n}})\id^{2}\hat{n}$, which in the tight-coupling approximation have negligible projections on to higher multipoles. The largest correction is from the quadrupole terms, but distortion anisotropies damp very fast so that this contribution vanishes rapidly.}.

In the tight coupling limit \citep{Hu1996anasmall}, the multipole hierarchy can be truncated after the quadrupole and corrections sourced by the damping of pure polarization terms may be neglected \citep{Chluba2015tens}. The contributions proportional to the monopole drop out of the final result, as well as the dipole parts after including second-order scattering terms \citep{Chluba2012}. Thus, only the quadrupole contribution is left, for which the relevant time derivative is given by $\dot{\Theta}_{2}\approx-(3/4)\,\dot{\tau}\,\Theta_{2}$ \citep{Hu1997}. The final result for the anisotropic heating rate thus is
\begin{align}\label{eq22}
\frac{\id}{\id t}\left[\frac{Q(\vek{x},t)}{\rho_{\gamma}}\right]
&\approx 15 \taudot \,\Theta^2_2(\vek{x},t)
\\[0mm]
\nonumber
&\!\!\!\!\!\!=15 \taudot \int 
\!\!\frac{\id^{3}k}{(2\pi)^3}\frac{\id^{3}k'}{(2\pi)^{3}}
\,\expf{i\vek{x}\cdot(\vek{k}+\vek{k}')}\,\mathcal{R}(\vek{k})\,\mathcal{R}(\vek{k}')
\,\Trans_2(k,t)\,\Trans_2(k',t),
\end{align}
where $\taudot=\sigT \Ne c$ is the time-derivative of the Thomson optical depth, $\Theta_\ell(\vek{x},t)=\frac{1}{4\pi} \int P_\ell(\vgh\cdot \vghp)\, \Theta(\vek{x},\vghp, t) \id^2 \vghp$ and $\Trans_\ell(k,t)$ denotes the radiation transfer functions in $k$-space. The latter only depends on $k=|\vek{k}|$ and map the initial curvature perturbation amplitude, $\mathcal{R}(\vek{k})$, into temperature anisotropies at a later stage. In the tight coupling regime ($z\gtrsim 10^3$), we have \citep{Hu1996anasmall}
\beal
\label{eq:T2_approx}
\Trans_2(k, t) &\approx \frac{8k D c}{15\taudot a}\frac{\sin (k\rs)}{\sqrt{3}}\,\expf{-k^2/\kD^2}
\end{align}
where $D=(1+4R_\nu/15)^{-1}$ with $R_\nu \approx 0.41$ for three massless neutrinos; $\rs\approx \eta /\sqrt{3}$ denotes the sound horizon; $\eta=\int c\id t/a$ is conformal time; $a=(1+z)^{-1}$ is the scale factor and $\kD$ is the standard photon damping scale, which in the radiation-dominated era is determined by $\kD^{-2}=\int_z^\infty \frac{8c^2}{45 \taudot  a H}\id z'$, or $\kD\approx \pot{4.0}{-6}(1+z)^{3/2}\,\Mpc^{-1}$.

For the final type of the distortion fluctuations, it is important {\it when} the energy is released. Here, we shall describe the final signal as a superposition of $\mu$ and $y$ distortions with an additional temperature shift. The $\mu$ and $y$ contributions can be modeled using energy branching ratios, $\mathcal{J}_{\mu}(z)$ and $\mathcal{J}_{y}(z)$, for the $\mu$ and $y$ distortions \citep{Chluba2013Green}. This approximation assumes that the distortion shape only depends on the total number of Compton scatterings but not on the anisotropies themselves. Different approximations for the distortion visibilities are summarized in \citet{Chluba2016}. Unless stated otherwise, here we will use \citep{Chluba2013Green}
\bsub
\label{eq:branching_approx_improved}
\begin{align}
\mathcal{J}_y(z)
&\approx
\begin{cases}
\left(1+\left[\frac{1+z}{\pot{6}{4}}\right]^{2.58}\right)^{-1}
 & \text{for}\; z_{\rm rec}\leq z 
\\
0 & \text{otherwise}
\end{cases}
\\[0mm]
\mathcal{J}_\mu(z) 
&\approx\mathcal{J}_{\rm bb}(z)\,\left[1-\exp\left(-\left[\frac{1+z}{\pot{5.8}{4}}\right]^{1.88}\right)\right],
\end{align}
\esub
where $z_{\rm rec}=10^3$ and $\mathcal{J}_{\rm bb}(z) = \exp\left[-(z/z_{\mu})^{5/2}\right]$ accounts for the effect of thermalization, which becomes very efficient at redshifts $z\gtrsim z_{\mu}=\pot{2}{6}$ and erases the distortions.

With these definitions, we can write the spatially varying distortions caused by the dissipation process as
\bsub
\label{eq:mu_y_spatial}
\begin{align}
\label{eq4}
y(\vek{x},z)&\approx \frac{1}{4} \int^{\infty}_{z} 
\frac{\id}{\id z'}\left[\frac{Q(\vek{x},z')}{\rho_{\gamma}}\right] \mathcal{J}_{y}(z') \id z'\,
\\[1mm]
\label{eq5}
\mu(\vek{x},z)&\approx 1.401 \int^{\infty}_{z} 
\frac{\id}{\id z'}\left[\frac{Q(\vek{x},z')}{\rho_{\gamma}}\right] \,\mathcal{J}_{\mu}(z')\id z',
\end{align}
\esub
where we used the redshift as the time coordinate\footnote{We define $\id [Q(\vek{x},z)/\rho_{\gamma}]/\id z>0$ for energy release.}. These expressions are very similar to those for uniform energy release \citep[e.g., Eq.~(6) in][]{Chluba2013PCA}. The only difference is that here the energy release rate varies spatially. This causes anisotropic $\mu$ and $y$ distortions with the degree of anisotropy depending on the curvature power spectrum at small scales. 

Similar to the uniform $\mu$ and $y$ distortions from the dissipation of acoustic modes, we can introduce $k$-space window functions that link the distortion and curvature perturbations \citep[see][]{Chluba2012inflaton, Chluba2013iso}. From Eqs. \eqref{eq22} and \eqref{eq:mu_y_spatial} we then have
\begin{equation}
\label{eq:X_def_gen}
X(\vek{x},z)\approx\int\!\! \frac{\id^{3}k}{(2\pi)^3}\frac{\id^{3}k'}{(2\pi)^{3}}
\,\expf{i\vek{x}\cdot(\vek{k}+\vek{k}')}\,\mathcal{R}(\vek{k})\,\mathcal{R}(\vek{k}') \, W_{X}(k,k',z),
\end{equation}
where $X\in\{\mu,y\}$. The $k$-space window functions capture all the thermalization physics and mode coupling. In compact form they can be rewritten as
\begin{align}
\label{eq:W_X}
W_{X}(k,k',z)
&= 15 \int_z^\infty \frac{\taudot a}{H} \,\mathcal{G}_{X}(z')\,\Trans_2(k,z')\,\Trans_2(k',z')\id z'
\\ \nonumber
&\approx \frac{32}{45}\int_{z}^{\infty}\!\!
\frac{c^2 D^{2}}{\dot{\tau}a H} \,\mathcal{G}_{X}(z')
\,2 kk' \!\sin(k \rs)\sin(k' \rs)\,\expf{-\frac{k^{2}+{k'}^2}{\kD^{2}(z')}} \id z' ,
\end{align}
with $\mathcal{G}_{\mu}(z)=1.401\,\mathcal{J}_{\mu}(z)$ and $\mathcal{G}_{y}(z)=(1/4)\,\mathcal{J}_{y}(z)$. 

When using the transfer function approximation, Eq.~\eqref{eq:T2_approx}, for the computation of the window function, Eq.~\eqref{eq:W_X}, the momentum integrals in Eq.~\eqref{eq:X_def_gen} are carried out at $k>k_{\rm cut}\simeq 0.1\,\Mpc^{-1}$ because horizon scale modes at $z\simeq z_{\rm rec}$ do not dissipate. Alternatively, one can use more elaborate expressions for the transfer function or full numerical results to capture the super-horizon evolution \citep{Chluba2012}. However, we find that for our purposes the simpler approximation usually is sufficient.

\begin{figure}
\centering
\includegraphics[width=0.98\columnwidth]{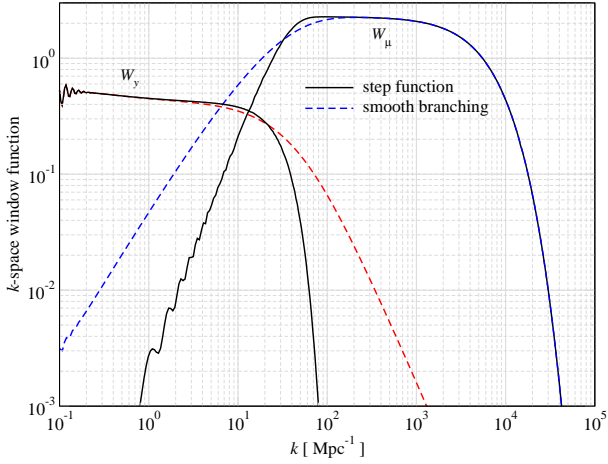}
\caption{Dependence of the $k$-space window functions, $W_X(k, k', z_{\rm rec})$, for $k=k'$ on the approximation for the energy branching ratios. Solid lines refer to the step-function approximation, while dashed lines show the window functions using $\mathcal{J}_X$ as in Eq.~\eqref{eq:branching_approx_improved}.}
\label{fig:W_mu_y_k_kp}
\end{figure}

\begin{figure}
\centering
\includegraphics[width=0.98\columnwidth]{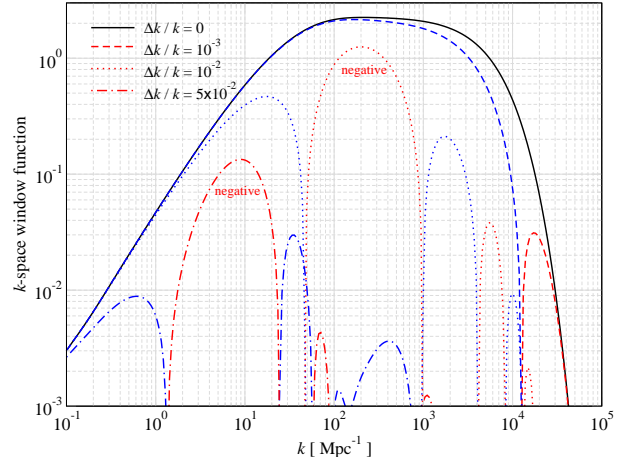}
\\[-0.5mm]
\includegraphics[width=0.98\columnwidth]{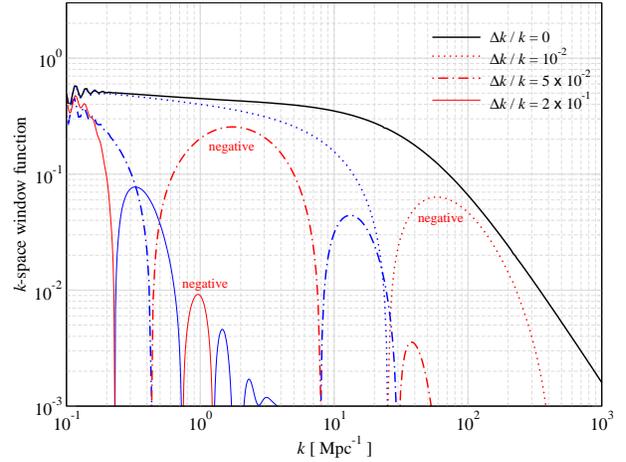}
\caption{Dependence of the $k$-space window functions, $W_X(k, k', z_{\rm rec})$, on the difference between $k$ and $k'$, where $\Delta k=k'-k$. The upper panel is for $\mu$, the lower for $y$. All red lines represent negative values.}
\label{fig:W_mu_y}
\end{figure}

\vspace{-2mm}
\subsection{Properties of the window functions, $W_{X}(k,k',z)$}
The window functions, $W_{X}(k,k',z)$, have a few simple properties. First, by symmetry $W_{X}(k,k',z)=W_{X}(k',k,z)$, so that one only has to consider cases $k\leq k'$. For $k=k'$, the window function reduces to the one for uniform heating, with no coupling between modes \citep{Chluba2012inflaton, Chluba2013iso}. The only difference is due to the approximations of the energy branching ratios, $\mathcal{J}_{X}(z)$. In earlier works, $\mathcal{J}_{y}(z)\approx \mathcal{H}(z-\zmuy)$ and $\mathcal{J}_{\mu}(z)\approx\mathcal{J}_{\rm bb}\,[1-\mathcal{H}(z-\zmuy)]$, where $\mathcal{H}(z-\zmuy)=1$ for $z<\zmuy=\pot{5}{4}$. The effect of this approximation is illustrated in Fig.~\ref{fig:W_mu_y_k_kp}. We set $z=z_{\rm rec}=10^3$, which is relevant to our discussion. Using the step-function approximation, both the $y$ and $\mu$ distortion window functions pick up contributions from a narrower range of scales. With the expressions in Eq.~\eqref{eq:branching_approx_improved}, the transition around $k\simeq 50\,\Mpc^{-1}$ is smoother. This is because modes of a given scale dissipate energy over a range of redshifts \citep[e.g., Fig.~1 in][]{Chluba2012inflaton}.

When considering non-Gaussianity, the mode coupling $k\neq k'$ also has to be included. Inspecting Eq.~\eqref{eq:W_X}, it becomes clear that for strongly disparate $k$ and $k'$ the window functions should decrease notably, mainly due to the exponential factor. For $k\approx k'$, one can also set $2\sin (k\rs) \sin(k'\rs)\approx 1$ without affecting the result for the window function significantly (it basically removes the small wiggles seen in Fig.~\ref{fig:W_mu_y_k_kp}). Similarly, for $k\neq k'$ one has 
$2\sin (k\rs) \sin(k'\rs)=\cos(\Delta k \rs)-\cos(k_+\rs) \approx \cos(\Delta k \rs)$, 
with the substitution $\Delta k=k'-k$ and $k_+=k+k'$. In the last step, we replaced terms that vary fast with time with their scale-averaged values. We confirmed that this approximation works extremely well, affecting the shapes of the window functions negligibly. However, this approximation eases the numerical computation significantly.

In Fig.~\ref{fig:W_mu_y}, we illustrate the numerical results for the $y$ and $\mu$ window functions for $k\neq k'$. For $W_\mu$, we found $W_\mu(k, k', z_{\rm rec})\approx W_\mu(k, k, z_{\rm rec})$ as long as $\Delta k/k\lesssim 10^{-3}$. For larger difference in the wavenumber, the amplitude of $W_\mu$ drops strongly. At $\Delta k/k\gtrsim 10^{-1}$ one can simply set $W_\mu\approx 0$ (confirmed numerically). Similarly, we have $W_y(k, k', z)\approx W_y(k, k, z)$ for $\Delta k/k\lesssim 10^{-2}$ and $W_y\approx 0$ for $\Delta k/k\gtrsim 0.5-1.0$ (confirmed numerically). This behavior of the $k$-space window functions eases the numerical computation of the distortion correlation functions significantly.

We also mention that while the heating rate for a single $k$-mode shows oscillatory behavior in time (which means that heating occurs at different phases of the wave propagation, when the spatial gradients in the temperature are largest), the window-function becomes quite smooth, representing the time-averaged heating rate for each mode. This point is often confused in the literature. In particular, the time (redshift) average is required and does not automatically drop out.

\vspace{-3mm}
\subsection{Free-streaming after recombination}
We neglected any spatial evolution of the distortion anisotropies, assuming that they are created {\it in situ} and remain there until decoupling at $z=z_{\rm rec}$. After decoupling we simply assume that the distortions free-stream to the observer across a distance $r_L\approx 14\,{\rm Gpc}$ to the last scattering surface. With the plane wave identity
\begin{align}
\label{eq:exp_identity}
\expf{i\vek{x}\cdot \vek{k}}
=4\pi\,\sum_{\ell}\sum_{m=-\ell}^{m=\ell} i^\ell j_\ell(k\,x)\,Y^\ast_{\ell m}(\hat{\vek{k}})\,Y_{\ell m}(\hat{\vek{x}}),
\end{align}
the spectral distortion anisotropies in $\mu$ and $y$ for an observer at the origin ($\vek{x}=-\hat{\vek{n}} r_L$) can thus be expanded into spherical harmonics, $Y_{\ell m}(\hat{\vek{n}})$, as
\begin{align}
\label{eq:a_X}
a_{\ell m}^{X}
&\equiv  \int X(\vek{x}, z_{\rm rec})\,Y^{*}_{\ell m}(\hat{\vek{n}}) \,\id^2{\hat n}
\nonumber\\[-1mm]
&=
4\pi\,(-i)^\ell \int\!\! \frac{\id^{3}k}{(2\pi)^3}\frac{\id^{3}k'}{(2\pi)^{3}} \,Y^{*}_{\ell m}(\hat{\vek{k}}_+)
\nonumber\\
&\qquad\qquad
\times j_\ell(k_+ r_L)\,\mathcal{R}(\vek{k})\,\mathcal{R}(\vek{k}') \, W_{X}(k,k',z_{\rm rec}),
\end{align}
where $\vek{k}_+=\vek{k}+\vek{k}'$. This expression can be used to compute approximate correlation functions between distortions and temperature anisotropies (see Sect.~\ref{correl_comput}). Corrections due to distortion transfer effects were neglected but will be discussed in Sect.~\ref{transfer_effects}.

\vspace{-3mm}
\subsubsection{Comparing with previous work}
To compare Eq.~\eqref{eq:a_X} for $a_{\ell m}^{\mu}$ directly with \citet{Pajer2012} and \citet{Ganc2012}, we approximate the window function, Eq.~\eqref{eq:W_X}. The first simplification is to replace the visibilities $\mathcal{J}_X(z)$ by step-functions. In this way, the redshift integral for the $\mu$-distortion is limited to $\zmuy \leq z \leq z_\mu$, where the redshift $\zmuy\approx \pot{5}{4}$ marks the transition between the $\mu$ and $y$-distortion eras \citep[e.g.,][]{Burigana1991, Hu1993}. If we then also replace the transfer function factor, $2\sin(k\rs)\sin(k'\rs)\approx 1$, we find 
\begin{align}
\label{eq:W_X_approx}
W_{\mu}(k,k',z)
&\approx 
\frac{32}{45}\int_{\zmuy}^{z_\mu} \frac{c^2 D^{2} \alpha_\mu}{\dot{\tau}a H}
\,kk' \expf{-\frac{k^{2}+{k'}^2}{\kD^{2}(z')}} \id z' 
\nonumber\\[-0.5mm]
&\approx 4\,D^{2}\alpha_\mu\,\frac{kk'}{k^{2}+{k'}^2} \int_{\zmuy}^{z_\mu} \partial_{z'} \expf{-(k^{2}+{k'}^2)/\kD^{2}(z')} \id z'
\nonumber\\[-1.5mm]
&\stackrel{\stackrel{k\approx k'}{\downarrow}}{\approx} 2\,D^{2} \alpha_\mu \left[\expf{-(k^{2}+{k'}^2)/\kD^{2}(z)}\right]_{\zmuy}^{z_\mu},
\end{align}
where $\alpha_\mu=1.401$ and $2\,D^{2}\alpha_\mu\approx 2.27$. Inserting this back into Eq.~\eqref{eq:a_X}, we obtain
\begin{align}
\label{eq:a_Xmu_Ganc}
a_{\ell m}^{\mu}
&\approx 
9.1\pi\,(-i)^\ell \int\!\! \frac{\id^{3}k}{(2\pi)^3}\frac{\id^{3}k'}{(2\pi)^{3}} \,Y^{*}_{\ell m}(\hat{\vek{k}}_+)
\nonumber\\
&\qquad\qquad
\times j_\ell(k_+ r_L)\,\mathcal{R}(\vek{k})\,\mathcal{R}(\vek{k}') 
\left[\expf{-(k^{2}+{k'}^2)/\kD^{2}(z)}\right]_{\zmuy}^{z_\mu}.
\end{align}
Comparing this with Eq.~(40) of \citet{Ganc2012}, we find agreement once we set\footnote{One should actually use $\cos(k\rs)\cos(k'\rs)\rightarrow \sin(k\rs)\sin(k'\rs)$.} $\left<\cos(k\rs)\cos(k'\rs)\right>_p=1/2$ in their expression. The only small difference is that, following \citet{Pajer2012}, a filter function was added to the $k$-space integral. In our approach, this filter function is directly related to $W_\mu(k, k', z_{\rm rec})$, which vanishes when $k$ and $k'$ differ significantly (see Fig.~\ref{fig:W_mu_y}), but was neglected in Eq.~\eqref{eq:a_Xmu_Ganc}. Our approximation also includes a factor of $3/4$ \citep{Chluba2012, Inogamov2015} with respect to \citet{Pajer2012}, which was based on the classical treatment of the dissipation problem \citep{Sunyaev1970diss, Daly1991, Hu1994}.

\vspace{-0mm}
\subsection{Cross-correlation of $\mu$ and $y$ with temperature} 
\label{correl_comput}
We now compute the correlation functions for different combinations of $\mu$, $y$ and temperature perturbations.
The temperature anisotropies seen by an observer at the origin can be expanded into spherical harmonics
\beal
\label{eq:a_T}
a_{\ell m}^{T}&\equiv \int \id^{2}n \,\frac{\Delta T(\hat{\vek{n}})}{T} Y^{*}_{\ell m}(\hat{\vek{n}})
\\
&\approx\frac{12\,\pi}{5}(-i)^{\ell}\int\frac{\id^{3}k}{(2\pi)^{3}}\,\mathcal{R}(\vek{k})\,\Delta_{\ell}(k)\,Y^{*}_{\ell m}(\hat{\vek{k}}),\nonumber
\end{align}
where $\Delta_{\ell}(k)\simeq j_{\ell}(k\,r_{L})/3$ is the radiation transfer function in the Sachs-Wolfe limit.\footnote{In contrast to \cite{Ganc2012} we use $\mathcal{R}$ instead of $\zeta$, with the convention $\Delta T/T=-\zeta/5 = \mathcal{R}/5$.} Alternatively, it is possible to directly use numerical results for the transfer functions at different values of $k$. This is expected to enhance the final correlation function, $C_\ell^{XT}$, at small scales with respect to the Sachs-Wolfe approximation but reduce the overall signal-to-noise ratio due to cancelation effects \citep[e.g.,][]{Ganc2012}.

Using $\langle \mathcal{R}_{\vek{k}}\mathcal{R}_{\vek{k}'}\rangle= (2\pi)^{3}\delta^{(3)}(\vek{k}+\vek{k}')\,P_{\mathcal{R}}(k)$, with curvature power spectrum $P_{\mathcal{R}}(k)=2\pi^{2}\Delta^{2}(k)/k^{3}$, the temperature power spectrum in the Sachs-Wolfe limit is given by
\begin{align}
\label{eq:C_ell_TT}
&C_{\ell}^{TT, \rm SW}\approx\frac{4\pi}{25}\int \frac{\id k_{T}}{k_{T}} 
j_{\ell}^{2}(k_{T} r_{L})\,\Delta^{2}\left(k_{T}\right)\approx \frac{2\pi}{25}\frac{\Delta^{2}_0}{\ell(\ell+1)},
\end{align}
where in the last step we assumed a scale-invariant power spectrum ($\nS=1$) with amplitude $\Delta^{2}_0$.

We use the primordial bispectrum in the squeezed limit to describe the scale-dependent\footnote{This convention leads to $f_{\rm NL}=-f_{\rm NL}^{\rm WMAP}$ when comparing to the WMAP convention for the bispectrum \citep{Komatsu2001}.}
\begin{align}
\langle \mathcal{R}_{\vek{k}}\mathcal{R}_{\vek{k}'}\mathcal{R}_{\vek{k}_{T}}\rangle
&\approx (2\pi)^{3}\delta^{(3)}(\vek{k}+\vek{k}'+\vek{k}_{T})\frac{12 \,f_{\rm NL}(k)}{5}
\, 
P_{\mathcal{R}}(k)P_{\mathcal{R}}(k_{T}).
\end{align}
Here, $k\approx k'$ was already used, but we confirmed that even more generally this limit for the bispectrum is sufficient.
After some algebra (see Appendix~\ref{app:aXT_explicit}), the spectral distortion-temperature correlation functions take the simple form
$\langle (a_{\ell m}^{X})^{*}a_{\ell'm'}^{T}\rangle=\delta_{\ell\ell'}\delta_{mm'}\,C_{\ell}^{X T}$,
where the correlation coefficients are given by
\begin{align}
\label{eq:C_ell_XT}
&C_{\ell}^{X T}\approx \frac{48\pi}{25}
\!\int \!\frac{\id k_{T}}{k_{T}}  \frac{\id k}{k} j_{\ell}^{2}(k_{T} r_{L}) \,\Delta^{2}\left(k_{T}\right) 
\nonumber\\[0mm]
&\qquad\qquad\qquad\qquad\qquad\times 
f_{\rm NL}(k) \,\Delta^{2}(k)  \,\bar{W}_{X}(k_T,k,z_{\rm rec})
\end{align}
and the azimuthally averaged $k$-space window function is 
\begin{align}
\label{eq:W_X_bar}
\bar{W}_{X}(k_T,k,z_{\rm rec})
&=
\frac{1}{2}\int_{|k-k_T|}^{k+k_T}  W_{X}(k,k_1,z_{\rm rec}) \frac{k_1\!\id k_1}{k \,k_T},
\end{align}
with $k_1=(k^2+k^2_T-2k k_T\mu)^{1/2}$. The averaged window function receives most of its contributions from $k_1\simeq k$, so that $k_T\ll k$ is preferred. Assuming $k_T \ll k$, one can indeed replace $\bar{W}_{X}(k_T,k, z)\approx W_{X}(k, k, z)\equiv W_{X}(k, z)$, which gives
\begin{align}
\label{eq:C_ell_XT_simp}
C_{\ell}^{X T}&\approx \frac{48\pi}{25}\!\int \!\!\frac{\id k}{k} \frac{\id k_{T}}{k_{T}}
j_{\ell}^{2}(k_{T} r_{L})\,\Delta^{2}\left(k_{T}\right) f_{\rm NL}(k) \,\Delta^{2}(k)\,W_X(k,z_{\rm rec})
\nonumber\\
&=
 12 \;
\frac{4\pi}{25}\!\int \!\!\frac{\id k_{T}}{k_{T}} j_{\ell}^{2}(k_{T} r_{L}) \Delta^{2}\left(k_{T}\right)
\times \int \!\!\frac{\id k}{k} 
f_{\rm NL}(k) \Delta^{2}(k) W_{X}(k, z_{\rm rec})
\nonumber\\
&=
 12\,C_\ell^{TT, \rm SW} \!\int \!\!\frac{\id k}{k} 
\,f_{\rm NL}(k) \,\Delta^{2}(k)\,W_{X}(k, z_{\rm rec})
\nonumber\\
&=
 12\,C_\ell^{TT, \rm SW}\,\mathcal{H}_X
\approx
 12\,C_\ell^{TT, \rm SW} f_{\rm NL}(k_X)\,\left<X\right>,
\end{align}
where in the last step we assumed that $f_{\rm NL}$ scales slowly with $k$ and can simply be evaluated at $k_X=\{7\,\Mpc^{-1}, 740\,\Mpc^{-1}\}$, respectively relevant to the $y$ and $\mu$ eras \citep[see discussion][]{Emami2015}. Explicitly, this means that 
\begin{align}
\label{eq:J_X}
\mathcal{H}_X&= \int \id \ln k
\,f_{\rm NL}(k) \,\Delta^{2}(k)\,W_{X}(k, z_{\rm rec})
\approx
f_{\rm NL}(k_X)\,\left<X\right>
\end{align}
was assumed. For scenarios with significant scale-dependence of $f_{\rm NL}$ around the distortion pivot scales, $k_X=\{7\,\Mpc^{-1}, 740\,\Mpc^{-1}\}$, this integral can be evaluated numerically after specifying $f_{\rm NL}(k)$ and $\Delta^{2}(k)$. Below we briefly discuss $f_{\rm NL}(k)\propto (k/k_X)^{n_{\rm NL}}$; however, a more general consideration requires a case-by-case study, which is beyond the scope of this paper.

The approximation $C_{\ell}^{X T}\approx 12\,C_\ell^{TT, \rm SW} f_{\rm NL}(k_X)\,\left<X\right>$ was given in \citet{Emami2015} and shows explicitly that the cross-correlation depends on the {\it average value} of the distortion parameters, 
\beal
\label{eq:average_X}
\left<X\right>=\int \id\ln k \,\Delta^{2}(k) W_{X}(k, z_{\rm rec}).
\end{align}
Equation~\eqref{eq:C_ell_XT_simp} also highlights that for $f_{\rm NL}>0$ the temperature perturbations and distortions are {\it correlated} at the largest scales. On the other hand, for the WMAP convention for the bispectrum, $f^{\rm WMAP}_{\rm NL}=-f_{\rm NL}>0$ implies that the temperature perturbations and distortions are {\it anti-correlated} at the largest scales. Including the full temperature transfer function, the cross power spectrum $C_\ell^{\mu T}$ is expected to change sign at $\ell \simeq 40$, as previously explained by \citet{Ganc2012}. The same statement applies to $C_\ell^{y T}$.

By extracting the data from Fig.~3 of \citet{Ganc2012} using the ADS {\tt Dexter} tool, we find the ratio $\rho(\ell)=C_{\ell}^{\mu T}/C_{\ell}^{\mu T, \rm SW}$ of the full radiative transfer result, $C_{\ell}^{\mu T}$, in comparison to the Sachs-Wolfe approximation, $C_{\ell}^{\mu T, \rm SW}$, within their computation to be well represented by
\begin{align}
\label{eq:C_ell_XT_simp_rho}
\rho(\ell) &\approx
1.08\,[1-0.022 \ell-\pot{1.72}{-4}\,\ell^2 
\nonumber \\
&\qquad +\pot{2.00}{-6}\,\ell^3-\pot{4.56}{-9}\,\ell^4]
\end{align}
at $2\leq \ell \leq 200$. Since the large-scale power spectrum parameters are well-known \citep{Planck2015params}, we can obtain the approximation
\begin{align}
\label{eq:C_ell_XT_simp_rho_improved}
C_{\ell}^{X T}&\approx 12\,C_\ell^{TT, \rm SW,1} \rho(\ell)\,f_{\rm NL}(k_X)\,\left<X\right>,
\end{align}
where $C_\ell^{TT, \rm SW,1}$ is given by Eq.~\eqref{eq:C_ell_TT} with $\nS=1$. We will show below that this approximation indeed works extremely well; however, some modifications are required when $f_{\rm NL}$ varies noticeably.

\vspace{-3mm}
\subsection{Correlations of $\mu$ and $y$} 
\label{mu_y_correl_comput}
We carry out the computation for the distortion correlation functions for the Gaussian and non-Gaussian contributions in Appendix~\ref{app:distortion_auto}. The Gaussian part is (negligibly) small with a quasi white-noise power spectrum until transfer effects become important. Here we focus on the result for the non-Gaussian contribution $\propto \mathcal{O}(f_{\rm NL}^2)$. Defining the power spectra, $\langle (a_{\ell m}^{X})^{*}a_{\ell'm'}^{Y}\rangle=\delta_{\ell\ell'} \delta_{mm'}\,C_\ell^{XY}$, and using $P(k)=2\pi^{2}\Delta^{2}(k)/k^{3}$, from Eq.~\eqref{eq:hhf} we find
\beal
\label{eq:Corr_XY_final}
C_\ell^{XY}&\approx
144  \;\frac{4\pi}{25}\int\!\! 
\frac{\id k}{k}\, j_{\ell}^2(k r_L)\, \Delta^2(k)\,
\nonumber\\
&\quad
\times \int \frac{\id k_1}{k_1}\frac{\id k_2}{k_2} 
 \Delta^2(k_1)\,\Delta^2(k_2) \, f_{\rm NL}(k_1) f_{\rm NL}(k_2) 
\nonumber\\
&\qquad\qquad
\times W_{X}(k_1,z_{\rm rec}) W_{Y}(k_2,z_{\rm rec})
\nonumber\\
&=
144 \,C_\ell^{TT, \rm SW}\, \mathcal{H}_X \, \mathcal{H}_Y,
\nonumber\\
&\approx
144 \,C_\ell^{TT, \rm SW}\,f_{\rm NL}(k_X)\,f_{\rm NL}(k_Y) \left<X\right>\left<Y\right>,
\end{align}
where in the last step we again used Eq.~\eqref{eq:J_X}. This approximation was given in \citet{Emami2015} and again explicitly illustrates how the correlation function depends on the average values, $\left<X\right>$ and $\left<Y\right>$. Radiative transfer effects slightly modify the $\ell$-dependence of the correlation functions, but these effects only become important at $\ell\gtrsim 200$ (see below) and are not further discussed here.

\vspace{3mm}
\subsection{Numerical results for the cross-power spectra}
To obtain the results for the distortion-temperature correlation function, we assume a power-law power spectrum $\Delta^{2}(k)=\Delta_{\rm p}^2\,(k/k_0)^{\nS-1}$ and non-Gaussianity $f_{\rm NL}(k)=f^{\rm p}_{\rm NL}\,(k/k_0)^{n_{f_{\rm NL}}}$ around pivot scale $k_0=0.05\,\Mpc^{-1}$. We set $\nS=0.9645$ and $\Delta_{\rm p}^2=\pot{2.21}{-9}$ \citep{Planck2013params, Planck2015params}. The integral over $k$ in Eq.~\eqref{eq:C_ell_XT} runs from $k_{\rm cut}=0.1\,\Mpc^{-1}$ to $k_{\rm max}=10^5\,\Mpc^{-1}$. This includes basically all modes that can lead to distortions in the $\mu$ and $y$-eras before recombination ends.

To ease the numerical calculation as a function of $\ell$, we define the function
\begin{align}
\label{eq:C_ell_XT_k}
&\frac{\id C^{X T}(k_T)}{\id\ln k_T}
=\int  \id \ln k\,f_{\rm NL}(k) \,\Delta^{2}(k)\, \bar{W}_{X}(k,k_T,z_{\rm rec}) 
\end{align}
such that $C_\ell^{X T}=12\int \frac{\id C^{X T}(k_T)}{\id \ln k_T}\,\frac{4\pi}{25}\,\Delta^{2}(k_T)\,j_\ell^2(k_T r_L) \id \ln k_T$. Thus, the $\ell$-dependence of the correlation function can be obtained after tabulating $\id C^{X T}(k_T)/\id\ln k_T$. For the $\mu-T$ correlation, we can furthermore directly compare with previous calculations \citep{Pajer2012, Ganc2012} by setting 
\begin{align}
\label{eq:W_Pajer}
\bar{W}_{\mu}(k,k_T,z_{\rm rec})\approx 2.27 \left[\expf{-2k^{2}/\kD^{2}(z)}\right]_{\zmuy}^{z_\mu} 
W_{\rm f}\left(k_T/\kD^{2}(\zmuy)\right),
\end{align}
where $W_{\rm f}(x)=3(\sin x- x\cos x)/x^3$ is a filter function that was introduced by hand.

\begin{figure}
\centering
\includegraphics[width=\columnwidth]{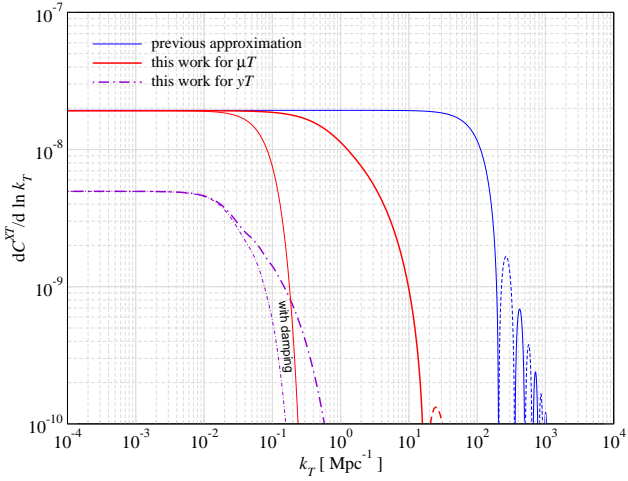}
\caption{Differential correlation function, $\id C^{X T}(k_T)/\id \ln k_T$, for constant $f_{\rm NL}(k)=1$. We compare Eq.~\eqref{eq:C_ell_XT_k} using the full $k$-space window function (red lines), Eq.~\eqref{eq:W_X} in Eq.~\eqref{eq:W_X_bar}, with the window-function approximation based on \citet{Pajer2012}, Eq.~\eqref{eq:W_Pajer} [blue line]. The purple line shows the result for the differential $y-T$ correlation. We also show the modification due to damping of distortions (thin lines), as discussed in Sect.~\ref{transfer_effects}.}
\label{fig:dC_XT}
\end{figure}
In Fig.~\ref{fig:dC_XT}, we show the differential correlation function, $\id C^{\mu T}(k_T)/\id \ln k_T$ for $f_{\rm NL}=1$, using our approach and the one of \citet{Pajer2012}. The approximation of \citet{Pajer2012} picks up contributions from significantly smaller scales. However, most of the total correlation arises from large scales, with $k_T\lesssim 0.01\,\Mpc^{-1}-0.1\,\Mpc^{-1}$. Thus, the final result for the $\mu-T$ correlation is hardly affected by the exact shape of the window function at $k_T\gtrsim 0.1\,\Mpc^{-1}$. In this case, one can indeed set $\bar{W}_{X}(k,k_T,z_{\rm rec})\approx W_{X}(k, z_{\rm rec})$, such that with
\begin{align}
&\frac{\id C^{\mu T}(k_T\rightarrow 0)}{\id\ln k_T}
\approx \mathcal{H}_\mu\approx f_{\rm NL}(k_\mu)\,\left<\mu\right>
\end{align}
the simple approximation $C_\ell^{\mu T}\approx 12\,C_\ell^{TT, \rm SW, 1} \rho(\ell)\, f_{\rm NL}(k_\mu)\,\left<\mu\right>$ works extremely well.  

In Fig.~\ref{fig:dC_XT}, we also show the differential correlation function for the $y-T$ correlation. Its amplitude at large scales is about four times lower than for the $\mu$ distortion, owing to the different relations of dissipated energy and distortion amplitude [i.e., $y\simeq (1/4) \Delta\rho_\gamma/\rho_\gamma$ versus $\mu\simeq 1.4 \Delta\rho_\gamma/\rho_\gamma$]. The $y-T$ correlation mainly picks up contributions from $k_T\lesssim 0.05\,\Mpc^{-1}$, so that the approximation $C_\ell^{y T}\approx 12\,C_\ell^{TT, \rm SW, 1} \rho(\ell)\,f_{\rm NL}(k_y)\,\left<y\right>$ is again well-justified.

\subsubsection{Simple formulae for varying $\nS$ and $n_{f_{\rm NL}}$}
For power-law dependence of $f_{\rm NL}$, with $n_{f_{\rm NL}}\neq 0$, one can capture the modifications by computing the average value $\left<\mu\right>$, Eq.~\eqref{eq:average_X}, with effective spectral index $\nS^*=\nS+n_{f_{\rm NL}}$ around modified pivot-scale $k_0=k_\mu\simeq 740\,\Mpc^{-1}$. We find
\begin{align}
\label{eq:C_ell_XT_simp_rho_generalization}
C_{\ell}^{\mu T}&\approx
12\,C_\ell^{TT, \rm SW, 1} \rho(\ell)\,f_{\rm NL}(k_\mu)\,\Delta^2(k_\mu) \, I_\mu(\nS^*)
\nonumber\\
\ln I_\mu(n)&\approx2.60\,[1-0.290 \xi+ 0.471\,\xi^2 -0.0516\,\xi^3]
\end{align}
with $\xi=n-1$ to represent the numerical result very well. The effective spectral index is $\nS^*=\nS^*(k_\mu)=\nS(k_\mu)+n_{f_{\rm NL}}(k_\mu)$, which can strongly differ from the values obtained at large angular scales. The expression for $\ln I_\mu(n)$ gives very similar results as the one presented in \citet{Chluba2013iso}; however, here we used the slightly improved distortion window-function, with distortion visibility according to Eq.~\eqref{eq:branching_approx_improved}, and shifted the pivot-scale to $k_\mu \simeq 740\,\Mpc^{-1}$. 

Similarly, for $C_{\ell}^{y T}$ we can write
\begin{align}
\label{eq:C_ell_yT_simp_rho_generalization}
C_{\ell}^{y T}&\approx
12\,C_\ell^{TT, \rm SW, 1} \rho(\ell)\,f_{\rm NL}(k_y)\,\Delta^2(k_y) \, I_y(\nS^*)
\nonumber\\
\ln I_y(n)&\approx0.81\,[1-1.44 \xi+ 2.39\,\xi^2 -0.214\,\xi^3]
\end{align}
with $\nS^*=\nS^*(k_y)=\nS(k_y)+n_{f_{\rm NL}}(k_y)$ and $k_y\simeq 7\,\Mpc^{-1}$.

The expressions given above clearly show that the temperature-distortion correlations can be enhanced in three main ways: i) by increasing $f_{\rm NL}(k_X)$, ii) by increasing the amplitude of the small-scale power spectrum, $\Delta^2(k_X)$, or iii) by changing the scaling of non-Gaussianty or the small-scale power spectrum around the distortion pivot scales, captured by $n_{\rm S}^*$. The first two effects affect the results most significantly, while the latter is less important unless $\nS^*$ deviates strongly from unity. 

\vspace{-0mm}
\subsection{Estimates for the observability of the $\mu-T$ correlation}
\label{sec:results_muT}
The $\ell$-dependence of the correlation functions in both cases is fully determined by $C_\ell^{TT, \rm SW,1} \rho(\ell)$. We can thus estimate the expected signal-to-noise ratio using \citep[e.g.,][]{Ganc2012}
\begin{align}
\label{eq:SN}
\left(\frac{S}{N}\right)^2&\approx
\sum_{\ell=2}^{\ell_{\rm max}} \frac{(2\ell+1) \left(C_{\ell}^{X T}\right)^2}{C_{\ell}^{TT}\,C_{\ell}^{XX,N}}
\nonumber\\
&=144\,\mathcal{H}_X^2\,
\sum_{\ell=2}^{\ell_{\rm max}} 
\frac{(2\ell+1) \left(C_\ell^{TT, \rm SW,1}\right)^2 \rho^2(\ell)}{C_{\ell}^{TT}\,C_{\ell}^{XX,N}},
\end{align}
where $C_{\ell}^{TT}$ is the CMB temperature power spectrum and $C_{\ell}^{XX,N}$ the noise level for the distortions. For {\it PIXIE} \citep[see][]{Pajer2012} we have $C_{\ell}^{XX,N}\approx 4\pi \, X^2_{\rm min}\,\expf{\ell^2/84^2}$, where $X_{\rm min}$ denotes the smallest detectable distortion monopole signal. 

\vspace{-0mm}
\subsubsection{Estimate in the Sachs-Wolfe limit}
To compare with previous results, we first obtain an estimate for the signal-to-noise ratio in the Sachs-Wolfe limit, which was used in several works \citep[e.g.,][]{Pajer2012, Emami2015}. For this we assume $C_{\ell}^{TT}\approx C_\ell^{TT, \rm SW}$ with $\nS=1$ and set $\rho(\ell)=1$ in Eq.~\eqref{eq:SN}. This means
\begin{align}
\label{eq:SN2_SW}
\left(\frac{S}{N}\right)^2&\approx
144\,\mathcal{H}_X^2\,
\left(\frac{\Delta_{\rm p}^2}{50\,X^2_{\rm min}}\,\right)\sum_{\ell=2}^{\ell_{\rm max}} 
\frac{(2\ell+1)}{\ell(\ell+1)}\,\expf{-\ell^2/84^2}.
\end{align}
Carrying out the sum up to $l_{\rm max}=200$, we obtain
\begin{align}
\label{eq:SN_result_SW}
\left(\frac{S}{N}\right)
&\approx
12\,\mathcal{H}_X\,\left[
\frac{0.37}{X_{\rm min}}\,\sqrt{\Delta_{\rm p}^2}\right]\quad \text{(Sachs-Wolfe limit)}.
\end{align}
If we now assume $\mathcal{H}_X\approx f_{\rm NL}(k_X)\,\left<X\right>$, and limit ourselves to the $\mu-T$ correlation, we obtain
\begin{align}
\label{eq:limit_SW}
f_{\rm NL}(k_\mu)&\lesssim
2900
\left[\frac{\Delta_{\rm p}^2}{\pot{2.2}{-9}}\right]^{-\frac{1}{2}}
\left[\frac{\mu_{\rm min}}{\pot{1.4}{-8}}\right]
\left[\frac{\left<\mu\right>}{\pot{2.3}{-8}}\right]^{-1}
\end{align}
for the upper limits on $f_{\rm NL}(k_\mu)$. 
The $y-T$ correlation limit can be obtained in a similar way. 
%
%
The large-scale power spectrum amplitude, $\Delta_{\rm p}^2$, is well-constrained and was fixed to the {\it Planck} value. For the minimal $\mu$-distortion we used, $\mu_{\rm min}\approx \pot{1.4}{-8}$, relevant to a {\it PIXIE}-type experiment \citep{Chluba2013PCA}.
Several estimates for the average $\mu$-distortion caused by the damping of acoustic waves at small scales have been given \citep{Chluba2011therm, Chluba2012, Dent2012, Pajer2012b, Khatri2013forecast, Chluba2013PCA, Cabass2016}. Here, we used  $\left<\mu\right>=\pot{(2.3\pm0.14)}{-8}$ \citep{Chluba2016}. Notice that within $\Lambda$CDM the cooling of baryons relative to photons \citep{Chluba2011therm} is expected to reduce the observed value for $\left<\mu\right>$ by $\Delta \mu\simeq \pot{-0.3}{-8}$ \citep{Chluba2016}. However, for the anisotropic $\mu$-distortion signals this modification should be neglected.

To compare with \citet{Emami2015}, which envisioned a spectrometer comparable to {\it PRISM} \citep{PRISM2013WPII}, we use $\Delta_{\rm p}^2=\pot{2.4}{-9}$, $\mu_{\rm min}=10^{-9}$ and $\left<\mu\right>=\pot{2}{-8}$ and obtain $f_{\rm NL}(k_\mu)\lesssim 230$, which is in very good agreement with their result. To compare with \citet{Pajer2012}, we set $\Delta_{\rm p}^2=\pot{2.4}{-9}$ and $\mu_{\rm min}=10^{-8}$. For the average $\mu$-parameter, \citet{Pajer2012} used $\left<\mu\right>\approx \pot{4.2}{-8}$, within their approximation, which yields $(S/N)\simeq \pot{0.9}{-3}\,f_{\rm NL}(k_\mu)$ or $f_{\rm NL}(k_\mu)\lesssim 1100$. The main difference to our estimate is due to the overestimation of $\left<\mu\right>$ within their computation, which neglected the factor of $3/4$ \citep{Chluba2012} and assumed $\nS=1$ for the power spectrum integral. 
Finally, to compare with \citet{Ganc2012}, we have to use $\Delta_{\rm p}^2=\pot{2.46}{-9}$, $\mu_{\rm min}=10^{-8}$ and\footnote{This follows from Eq.~(53) of \citet{Ganc2012}.} $\left<\mu\right>=\pot{3}{-8}$, which implies $f_{\rm NL}(k_\mu)\lesssim 1500$. This is a little tighter than what follows from Fig.~4 in their paper, which gives $(S/N)\simeq \pot{0.58}{-3}\,f_{\rm NL}(k_\mu)$ or $f_{\rm NL}(k_\mu)\lesssim 1700$. The difference can be explained when explicitly computing $C_\ell^{TT, \rm SW}$ with $\nS=0.96$, which gives a slightly lower signal-to-noise ratio, $\left(S/N\right)
\approx12\,\mathcal{H}_X\,\frac{0.34}{X_{\rm min}}\,\sqrt{\Delta_{\rm p}^2}$. 
Overall this shows that with Eq.~\eqref{eq:limit_SW} we can obtain reliable estimates for different experimental  sensitivities and in the Sachs-Wolfe limit.

\vspace{-0mm}
\subsubsection{Including transfer effects for the temperature fluctuations}
\label{sec:trans_T_inc}
We now include transfer effects for the estimate. This means we explicitly use the {\it Planck} temperature power spectrum and also $\rho(\ell)\neq 1$ in Eq.~\eqref{eq:SN}. For the {\it Planck} cosmology \citep{Planck2015params}, we then find
\begin{align}
\label{eq:SN_result}
\left(\frac{S}{N}\right)
&\approx
12\,\mathcal{H}_X\,\left[
\frac{0.24}{X_{\rm min}}\,\sqrt{\Delta_{\rm p}^2}\right].
\end{align}
The value of the sum in Eq.~\eqref{eq:SN_result_SW} is overestimated by a factor of $\simeq 1.6$. Inserting $\mathcal{H}_X\approx f_{\rm NL}(k_X)\,\left<X\right>$ for $\mu$, we then obtain
\begin{align}
\label{eq:limits}
f_{\rm NL}(k_\mu)&\lesssim
4500
\left[\frac{\Delta_{\rm p}^2}{\pot{2.2}{-9}}\right]^{-\frac{1}{2}}
\!\left[\frac{\mu_{\rm min}}{\pot{1.4}{-8}}\right]
\left[\frac{\left<\mu\right>}{\pot{2.3}{-8}}\right]^{-1}
\end{align}
for the upper limits on $f_{\rm NL}(k_\mu)$. For a spectrometer comparable to {\it PRISM} \citep{PRISM2013WPII} with $\mu_{\rm min}\simeq 10^{-9}$, we would expect $f_{\rm NL}(k_\mu)\lesssim 320$. Using the numbers from \citet{Ganc2012}, we find $f_{\rm NL}(k_\mu)\lesssim 2400$, which is in good agreement with their estimate. However, the average value for $\mu$ was overestimated $\simeq 30\%$ and a slightly more optimistic value for the smallest observable $\mu$-distortion was assumed. Their limit is thus a factor of $\simeq 2$ tighter than ours. Overall, transfer effects for the CMB temperature anisotropies weaken the constraint by a factor of $\simeq 1.6$ over the simple Sachs-Wolfe estimate, Eq.~\eqref{eq:limit_SW}.

\vspace{-0mm}
\section{Anisotropic spectral distortions: inclusion of transfer effects} 
\label{transfer_effects}
We now give a simplified formulation of the radiative transfer problem which captures the damping of distortion perturbations but was neglected above. At first order perturbation theory we can write the evolution equation for the photon occupation number, $n(t, x, \vek{r}, \vgh)\simeq n^{(0)}(t, x, \vek{r}, \vgh)+n^{(1)}(t, x, \vek{r}, \vgh)$, at time $t$, location $\vek{r}$, in the direction $\vgh$ and at frequency $x=h\nu/k\Tg={\rm const}$, with CMB monopole temperature $\Tg(z)=T_0(1+z)$, as \citep[e.g.,][]{DodelsonBook}
\begin{align}
\label{eq:evol_0}
\frac{\partial n^{(1)}}{\partial t}+\frac{c\vgh}{a}\cdot \nabla n^{(1)}-x\frac{\partial n^{(0)}}{\partial x} \left(\frac{\partial \Phi}{\partial t}+ \frac{c\vgh}{a}\cdot \nabla\Psi \right)=\mathcal{C}[n].
\end{align}
Here, $n^{(0)}$ is the average CMB occupation number, $\Phi$ and $\Psi$ are the potential perturbation in conformal Newtonian gauge, and $\mathcal{C}[n]$ denotes the rather complicated collision term, that accounts for the effect of Thomson scattering and thermalization processes. 

We now assume that the distortions of the CMB are described by a simple superposition between $\mu$- and $y$-distortion with a temperature shift
\begin{align}
\nonumber
n^{(1)}(t, x, \vek{r}, \vgh)\approx G(x) \, \Theta(t, \vek{r}, \vgh)+Y(x) \, y(t, \vek{r}, \vgh) + M(x)\, \mu(t, \vek{r}, \vgh).
\end{align}
Here, the function $G(x)=-x\,\partial_x n^{(0)}=x\expf{x}/[\expf{x}-1]^2$ gives the temperature shift, $Y(x)=x^{-2}\partial_x x^4 \partial_x n^{(0)}=G(x)[x\coth(x/2)-4]$ defines the $y$-distortion, and $M(x)=G(x)[1/\beta -1/x]$ is the $\mu$-distortion with $\beta =3\zeta(3)/\zeta(2)\simeq 2.192$.

As shown by \citet{Chluba2012}, the photon mixing process, mediated by Thomson scattering, causes a local $y$-distortion source term in second order of the temperature perturbation plus a small source term which directly increases the local CMB monopole temperature (which we neglect here). In this approximation, the evolution of the photon temperature field fully decouples from the evolution of the distortions. Thermalization processes convert $y$-distortions into $\mu$-distortions until energy exchange via Compton scattering becomes inefficient; $\mu$-distortions are erased, decaying to a temperature shift until photon production ceases. The scattering induced conversion of the $y\rightarrow \mu\rightarrow \Theta$ can be captured with the distortion visibility functions, Eq.~\eqref{eq:branching_approx_improved}, which defines the effective source function for $y$- and $\mu$-distortions.

Assuming that distortion sources are only affecting the local monopole of $y$- and $\mu$-parameters, by sorting terms according to their spectral dependence, from Eq.~\eqref{eq:evol_0} we then find
\bsub
\label{eq:evol_0_Tymu}
\begin{align}
\label{eq:evol_0_Tymu_a}
\frac{\partial \Theta}{\partial t}+\frac{c\vgh}{a}\cdot \nabla \Theta
&\approx \!-\frac{\partial \Phi}{\partial t} \!-\! \frac{c\vgh}{a}\cdot \nabla\Psi 
\!+\!\taudot \left(\Theta_0+\frac{\Theta_2}{10}-\Theta + \beta \chi\right)
\\
\label{eq:evol_0_Tymu_b}
\frac{\partial X}{\partial t}+\frac{c\vgh}{a}\cdot \nabla X
&\approx \taudot \,S_X(t, \vek{r}) 
+\taudot \left(X_0+\frac{X_2}{10}-X \right),
\end{align}
\esub
where $X=\{y,\mu\}$, $\taudot=c\sigT\Ne$ is the time derivative of the Thomson optical depth, $\beta=\varv /c$ denotes the baryon speed and $\chi$ the direction cosine of the baryon velocity with respect to $\vgh$.
We neglected polarization effects and introduced $X_\ell(t, \vek{r}, \vgh) = \sum_{m=-\ell}^\ell X_{\ell m}(t, \vek{r})\,Y_{\ell m}(\vgh)$, with  the spherical harmonic coefficient, $X_{\ell m}=[X]_{\ell m}$, of $X(t, \vek{r}, \vgh)$.

Equation~\eqref{eq:evol_0_Tymu_a} describes the normal evolution of the photon temperature anisotropies including Thomson scattering and first order Doppler shifts.
Equation~\eqref{eq:evol_0_Tymu_b} describes the evolution of $\mu$- and $y$-parameters, with the two source functions $S_y$ and $S_\mu$, which read
\beal
\label{eq:S_ymu}
S_X(t, \vek{r})\approx \mathcal{G}_{X}(z) \,
\frac{\id}{\id \tau}\left[\frac{Q(t, \vek{r})}{\rho_\gamma}\right],
\end{align}
where the spatially varying heating rate is given by Eq.~\eqref{eq22}.

There is a big difference between the two equations. For temperature perturbations, the decay of the potential terms close to horizon-crossing leads to an enhancement of the temperature fluctuations. In addition, the presence of the Doppler-term, $\propto \beta \chi$, allows the formation of {\it pressure waves}, which are absent for the distortion field. The reason is that in the tight-coupling limit $3\Theta_1 \approx \beta \chi$, so that the evolution equation for the dipole amplitude remains largely undamped until baryons start slipping. At early times, the damping of temperature perturbations is thus mediated through the quadrupole anisotropy \citep[e.g.,][]{Hu1996anasmall}.

In contrast to the temperature evolution, the $\mu$-dipole is heavily damped, so that no propagating waves appear. It is therefore sufficient to follow modes up to $\ell=1$. The $\mu$-hierarchy in Fourier-space then reads \citep[see also][]{Pajer2012b}
\bsub
\label{eq:mu_equations}
\beal
\partial_\eta\mu_0+k\mu_1&\approx 0
\\
\partial_\eta\mu_1-\frac{k}{3}\mu_0&\approx -\tau' \mu_1,
\end{align}
\esub
with conformal time $\eta=\int \frac{c\id t}{a}$ and $\tau'=a \taudot/c$, which combines to
\beal
\label{eq:T0_equations}
\mu''_0+ \tau' \mu'_0+\frac{k^2}{3}\mu_0\approx 0,
\end{align}
where primes denote derivatives with respect to $\eta$. In the regime of interest, the solution is an over-damped oscillator (the second derivative can be neglected), so that 
\bsub
\label{eq:mu_solution}
\beal
\mu_0(\eta_i, \eta, k)&\approx \mu_0(\eta_i, k) \, \expf{-\frac{15}{8}\,k^2/\kD^2(\eta_i,\eta)}
\\
\mu_1(\eta_i, \eta, k)&\approx \frac{k}{3\tau'}\,\mu_0(\eta_i, \eta, k)
\end{align}
\esub
between $\eta_i$ and $\eta$, with $\kD^{-2}(\eta', \eta)=\kD^{-2}(\eta)-\kD^{-2}(\eta')$. This shows that distortion dipole terms are strongly suppressed until after decoupling and that higher anisotropies can be omitted. Also, the damping scale is about $\sqrt{15/8}\simeq 1.4$ times larger than for the temperature case, with $k_{\rm D, \mu}=\sqrt{8/15}\,\kD<\kD$. In real-space, this damping creates a simple Gaussian-smearing. Including a non-vanishing monopole source term we find
\beal
\label{eq:source_III}
\mu_0(\eta, k)\approx \int_0^\eta \left[\frac{[\tau' S_\mu(\eta', k)]'}{\tau'}+\tau' S_\mu(\eta', k)\right] \,\expf{-\frac{15}{8}\frac{k^2}{\kD^2(\eta', \eta)}}\id \eta',
\end{align}
where one can furthermore neglect the first contribution in the source integral. Using Eq.~\eqref{eq22}, we can compute the distortion source term in Fourier-space. Inserting this into Eq.~\eqref{eq:source_III}, we find
\beal
\label{eq:source_IV}
X_0(z, \vek{k})&\approx 
\!\!\int\!\!\frac{\id^3 k'}{(2\pi)^3}
\mathcal{R}(\vek{k}')\,\mathcal{R}(\vek{k}-\vek{k}') W^{\rm D}_X(k, k',|\vek{k}-\vek{k}'|,z)
\\ \nonumber
W^{\rm D}_X(k,k', k'',z)&\!=\!
15 \!\int_z^\infty \!\frac{\taudot a}{H} \,\mathcal{G}_{X}(z')\,\Trans_2(k',z')\,\Trans_2(k'',z')\,\expf{-\frac{15}{8}\frac{k^2}{\kD^2(z', z)}} \!\id z'
\end{align}
after transforming back to redshift. Thus, in real-space we again obtain Eq.~\eqref{eq:X_def_gen} but with $W_X(k, k', z)\rightarrow W^{\rm D}_X(|\vek{k}+\vek{k}'|,k, k',z)$ to include the damping of power by photon diffusion. Similarly, in Eq.~\eqref{eq:a_X} we have the replacement $W_X(k, k', z_{\rm rec})\rightarrow W^{\rm D}_X(k_+,k, k',z_{\rm rec})$. For the correlation functions, we only need to modify the expressions for $\bar{W}_X$, which is achieved by $W_{X}(k,k_1,z_{\rm rec})\rightarrow W^{\rm D}_X(k_T, k, k_1,z_{\rm rec})$ in Eq.~\eqref{eq:W_X_bar}, and similarly in Eq.~\eqref{eq:Corr_XY_final}. For the average distortion parameter, Eq.~\eqref{eq:average_X}, no change is needed. 
For numerical evaluations, it is sufficient to use 
\beal
\label{eq:source_V}
W^{\rm D}_X(k,k', k'',z)&\approx \expf{-\frac{15}{8}\frac{k^2}{\kD^2(z)}}\,W_X(k', k'',z).
\end{align}
The error that is introduced by this approximation only affects the window-function at the largest scales ($k\simeq 0.1\,\Mpc^{-1}$). For $\mu$-distortion anisotropies, this does not have any noticeable effect. This is because between the redshift when a given $k$-mode dissipates most of its energy, $z_{\rm diss}\approx \pot{4.5}{5}\,[k/10^3\,\Mpc^{-1}]^{2/3}$ \citep{Chluba2012inflaton} and $z_{\rm rec}\approx 10^3$, one has $\kD^{-2}(z', z)\approx \kD^{-2}(z)$, without any large error. This approximation is not as well justified for the $y$-distortion anisotropies, since a non-negligible contribution arises from modes with $k\simeq 0.1\,\Mpc$ (see Fig.~\ref{fig:W_mu_y}). Thus, the approximation slightly underestimates the total $y$-distortion anisotropies. 
With these assumptions, we finally have
\begin{align}
\label{eq:C_ell_XT_k_damped}
\frac{\id C^{X T, \rm D}(k_T)}{\id\ln k_T}
&\approx \expf{-\frac{k_T^2}{k_{\rm D, \mu}^2(z_{\rm rec})}}\,\frac{\id C^{X T}(k_T)}{\id\ln k_T}
\nonumber\\[-0.5mm]
\frac{\id C^{X Y, \rm D}(k_T)}{\id\ln k_T}
&\approx \expf{-\frac{2 k_T^2}{k_{\rm D, \mu}^2(z_{\rm rec})}}\,\frac{\id C^{X Y}(k_T)}{\id\ln k_T}
\end{align}
for the damped case with\footnote{We used $\kD(z_{\rm rec})=0.144\,\Mpc^{-1}$ for $z_{\rm rec}=1100$.} $k_{\rm D, \mu}(z_{\rm rec})=\sqrt{\frac{8}{15}}\,\kD(z_{\rm rec})\approx 0.11 \,\Mpc^{-1}$. This result is in good agreement with that from the arguments in \citet{Pajer2012b}. In particular, for the $\mu$-distortion correlations, we expect this to work very well, assuming that recombination is instantaneous. It also renders the differences for $\frac{\id C^{X T}(k_T)}{\id\ln k_T}$ at small scales (compare Fig.~\ref{fig:dC_XT}) irrelevant. This implies that we can furthermore evaluate the differential power spectra at the largest scales and then simplify the computation to
\begin{align}
C^{X T, \rm D}
&\approx 12 \frac{\id C^{X T}(k_T\rightarrow 0)}{\id\ln k_T}
\times \int \!\!\frac{4\pi}{25}\,\frac{\id k}{k} 
j_{\ell}^{2}(k r_{L}) \, \Delta^{2}\left(k\right) \,\expf{-\frac{k^2}{k_{\rm D, \mu}^2(z_{\rm rec})}}
\nonumber\\ \nonumber
C^{X Y, \rm D}
&\approx 144\frac{\id C^{X Y}(k_T\rightarrow0)}{\id\ln k_T} \times \int \!\!\frac{4\pi}{25}\,\frac{\id k}{k} 
j_{\ell}^{2}(k r_{L}) \,\Delta^{2}\left(k\right) \,\expf{-\frac{2 k^2}{k_{\rm D, \mu}^2(z_{\rm rec})}}\,.
\end{align}
For higher accuracy, one would replace $j_{\ell}^{2}(k r_{L})\rightarrow 3 \Delta_\ell(k) \,j_{\ell}(k r_{L})$ in $C^{X T, \rm D}$ to correctly include transfer effects for the temperature anisotropy part. We explicitly confirmed that the effect of damping only becomes noticeable at $\ell \gtrsim 200$. This leaves previous signal-to-noise ratio estimates for measurements of $\mu$-$T$ correlations largely unaffected and can be neglected.

\vspace{-0mm}
\section{Conclusions}
\label{sec:conclusions}
We provided a step-by-step formulation for the evolution of distortion anisotropies created by the damping of small-scale acoustic modes in the early Universe. We include a treatment for the transition between $\mu$ and $y$-distortions as well as an approximate solution for the damping of distortion anisotropies. The calculations can be greatly simplified using Fourier-space window functions that capture the effects of mode-coupling and thermalization. The suggested approach could also be useful when developing a more detailed real-space radiative transfer treatment, which goes beyond the tight-coupling limit.

In Sect.~\ref{sec:results_muT}, we compare estimates for the observability of the $\mu-T$ correlation in the presence of non-Gaussianity previously given in the literature, clarifying the origin of their differences (see Sect.~\ref{sec:results_muT} for  discussion). A reliable estimate can be obtained with Eq.~\eqref{eq:limits}, yielding an upper limit of $f_{\rm NL}(k_\mu)\lesssim 4500\,\left[\frac{\left<\mu\right>}{\pot{2.3}{-8}}\right]^{-1}$ ($68\%$ c.l.) for experimental parameters similar to {\it PIXIE} should no $\mu$-distortion anisotropy be seen. This probes $f_{\rm NL}$ at scales with wavenumber $k_\mu\simeq 740\,\Mpc^{-1}$, which is at much smaller scales than constraints on non-Gaussianity derived using the CMB temperature anisotropies \citep{Planck2013ng}. 
Transfer effects related to the CMB temperature anisotropies (see Sect.~\ref{sec:trans_T_inc}) weaken the limit by a factor of $\simeq 1.6$ relative to what is obtained in the Sachs-Wolfe approximation, Eq.~\eqref{eq:limit_SW}.

Since the derived upper limit depends inversely on the average $\mu$-distortion that is created by the dissipation process, strongly enhanced small-scale power could render distortion anisotropies observable for much smaller $f_{\rm NL}$. To break the degeneracy, an absolute measurement of the CMB monopole distortion has to be performed. This shows that there is a fundamental difference between CMB imaging concepts (e.g., {\it CORE, Litebird}) and spectrometers like {\it PIXIE} when it comes to interpreting the constraints. 

Overall, spectral distortions provide an interesting new avenue for testing non-Gaussianity at scales that are inaccessible by other means. It will be important to explore how different sources of distortions (and their anisotropies) could be distinguished, as the average value of $\mu$ can in principle be changed in non-standard scenarios \citep[e.g.,][]{Sarkar1984, Hu1993b, McDonald2001, Chluba2011therm, Tashiro2012, Chluba2013fore, Chluba2013PCA, Tashiro2013, Amin2014, Yacine2015, Ema2016G}. It will furthermore be crucial to carry out realistic forecasts including real-world limitations caused by foregrounds. 

\small

\section*{Acknowledgments}
The authors cordially thank the referee for helpful comments on the manuscript. They furthermore thank Eiichiro Komatsu for helpful discussions about the definition of $f_{\rm NL}$ with respect to the commonly used WMAP parametrization.
JC is supported by the Royal Society as a Royal Society URF at the University of Manchester, UK. 
ED acknowledges support from the DOE under grant No. de-sc0008016. MK was supported by NSF Grant No. 0244990, NASA NNX15AB18G, and the Simons Foundation.

\bibliographystyle{mn2e}
\bibliography{Lit}

\begin{appendix}

\section{Local-model non-Gaussian curvature perturbations}
\label{app:non-Gaussian}
The local-model scale-independent non-Gaussian curvature perturbations in real-space may be written as
\beal
\mathcal{R}(\vek{x})=\mathcal{R}^{\rm G}(\vek{x})+\frac{3}{5}\,f_{\rm NL} \,[\mathcal{R}^{\rm G}(\vek{x})]^2,
\end{align}
where $\mathcal{R}^{\rm G}(\vek{x})$ is a Gaussian random field. With this definition, we have 
\beal
\langle\mathcal{R}(\vek{x})\rangle=\frac{3}{5}\,f_{\rm NL} \int \frac{\id^3 q}{(2\pi)^3}P(q)
=\frac{3}{5}\,f_{\rm NL}\,\mathcal{I},
\end{align}
where we transformed to Fourier-space and then used $\langle \mathcal{R}^{\rm G}(\vek{k})\rangle=0$ and $\langle \mathcal{R}^{\rm G}(\vek{k})\mathcal{R}^{\rm G}(\vek{k'})\rangle=(2\pi)^3\,\delta^{(3)}(\vek{k}+\vek{k}')\,P(k)$. Here, $P(k)$ is the power spectrum of the Gaussian random field. Formally, the integral $\mathcal{I}$ is not well-defined and physical quantities should not depend on its value. We thus use
\beal
\mathcal{R}(\vek{x})=\mathcal{R}^{\rm G}(\vek{x})+\frac{3}{5}\,f_{\rm NL}\left([\mathcal{R}^{\rm G}(\vek{x})]^2-\mathcal{I}\right),
\end{align}
instead of the initial expression. Transforming into Fourier-space, we then find the ensemble averages for $\mathcal{R}_i=\mathcal{R}(k_i)=\mathcal{R}_i^{\rm G}+\Delta \mathcal{R}_i$
\bsub
\beal
\langle \mathcal{R}_1\,\mathcal{R}_2\rangle
&=(2\pi)^3\,\delta^{(3)}(\vek{k}_1+\vek{k}_2)\,P(k_1) 
\\ \nonumber
%
&\qquad +(2\pi)^3\,\delta^{(3)}(\vek{k}_1+\vek{k}_2)\,
\frac{18}{25}\,f_{\rm NL}^2 \int \frac{\id^3 q}{(2\pi)^3}P(q)\,P(|\vek{k}_1-\vek{q}|)
\\
\langle \mathcal{R}_1\,\mathcal{R}_2\,\mathcal{R}_3\rangle
&=
(2\pi)^3\,\delta^{(3)}(\vek{k}_1+\vek{k}_2+\vek{k}_3)
\\ \nonumber
&\qquad\times\frac{6}{5}\,f_{\rm NL}\,\left[P(k_1)\,P(k_2) + P(k_1)\,P(k_3)+P(k_2)\,P(k_3)\right]
\\
\langle \mathcal{R}_1\,\mathcal{R}_2\,\mathcal{R}_3\,\mathcal{R}_4\rangle
&=\langle \mathcal{R}^{\rm G}_1\,\mathcal{R}^{\rm G}_2\,\mathcal{R}^{\rm G}_3\,\mathcal{R}^{\rm G}_4\rangle
+(2\pi)^3\,\delta^{(3)}(\vek{k}_1+\vek{k}_2+\vek{k}_3+\vek{k}_4)
\\ \nonumber
&\!\!\!\!\times\frac{36}{25}\,f^2_{\rm NL}\,\left\{P(k_1)\,P(k_2)[P(k_{13})+P(k_{14})] + 5\, {\rm permutations} \right\}.
\end{align}
\esub
for scale-independent $f_{\rm NL}$. Here, $\vek{k}_{ij}=\vek{k}_i+\vek{k}_j$.
For the scale-dependent case, we similarly find
\bsub
\beal
\langle \mathcal{R}_1\,\mathcal{R}_2\rangle
&=(2\pi)^3\,\delta^{(3)}(\vek{k}_1+\vek{k}_2)\,P(k_1) 
\\ \nonumber
&+(2\pi)^3\,\delta^{(3)}(\vek{k}_1+\vek{k}_2)\,
\frac{18}{25}\,f_{\rm NL}(k_1)^2 \!\int \!\frac{\!\id^3 q}{(2\pi)^3}P(q)\,P(|\vek{k}_1-\vek{q}|)
\\
\langle \mathcal{R}_1\,\mathcal{R}_2\,\mathcal{R}_3\rangle
&=
(2\pi)^3\,\delta^{(3)}(\vek{k}_1+\vek{k}_2+\vek{k}_3)
\nonumber\\
&\qquad\times\frac{6}{5}\,\left[P(k_1)\,P(k_2) \,f_{\rm NL}(k_3)+ 2\, {\rm permutations}\right]
\\
\label{eq:R4_nonG}
\langle \mathcal{R}_1\,\mathcal{R}_2\,\mathcal{R}_3\,\mathcal{R}_4\rangle
&=\langle \mathcal{R}^{\rm G}_1\,\mathcal{R}^{\rm G}_2\,\mathcal{R}^{\rm G}_3\,\mathcal{R}^{\rm G}_4\rangle
+(2\pi)^3\,\delta^{(3)}(\vek{k}_1+\vek{k}_2+\vek{k}_3+\vek{k}_4)
\\ \nonumber
&\!\!\!\!\!\!\!\!\!\!\!\!\!\!\!\!\!\!\!\!\!\!\!\!\!\!\!\!
\times\frac{36}{25}\left\{P(k_1)\,P(k_2)[P(k_{13})+P(k_{14})] f_{\rm NL}(k_3) f_{\rm NL}(k_4)+ 5\, {\rm permutations} \right\}.
\end{align}
\esub
The trispectrum contributions are $\mathcal{O}(\Delta^6)$, so unless $f_{\rm NL}$ is significantly larger than unity, this contribution is negligible relative to the Gaussian part.

\newpage
\section{Distortion-temperature correlation function}
\label{app:aXT_explicit}
Using Eq.~\eqref{eq:a_X} and \eqref{eq:a_T}, we can compute the distortion-temperature correlation function as 
\beal
\langle (a_{\ell m}^{X})^{*}a_{\ell'm'}^{T}\rangle&=\langle a_{\ell m}^{X} (a_{\ell'm'}^{T})^{*}\rangle
\nonumber\\
&=
\frac{(4\pi)^2}{5}\,(-i)^\ell i^{\ell'} \int\!\! \frac{\id^{3}k_T}{(2\pi)^3}\frac{\id^{3}k}{(2\pi)^3}\frac{\id^{3}k'}{(2\pi)^{3}} \,
Y^{*}_{\ell m}(\hat{\vek{k}}_+)\,Y_{\ell' m'}(\hat{\vek{k}}_T)
\nonumber\\
&\qquad\qquad
\times j_{\ell}(k_+ r_L)\,j_{\ell'}(k_T r_L)
\nonumber\\
&\qquad\qquad
\times \langle \mathcal{R}(-\vek{k}_T) \mathcal{R}(\vek{k})\,\mathcal{R}(\vek{k}')\rangle \, W_{X}(k,k',z_{\rm rec}),
\end{align}
with $\vek{k}_+=\vek{k}+\vek{k}'$. Assuming scale-independent $f_{\rm NL}$, we then have
\beal
\langle (a_{\ell m}^{X})^{*}a_{\ell'm'}^{T}\rangle&=
\frac{(4\pi)^2}{5}\,(-i)^\ell i^{\ell'} \int\!\! \frac{\id^{3}k_T}{(2\pi)^3}\frac{\id^{3}k}{(2\pi)^3}\frac{\id^{3}k'}{(2\pi)^{3}} \,
Y^{*}_{\ell m}(\hat{\vek{k}}_+)\,Y_{\ell' m'}(\hat{\vek{k}}_T)
\nonumber\\
&\qquad\qquad
\times j_{\ell}(k_+ r_L)\,j_{\ell'}(k_T r_L)
\nonumber\\
&\qquad\qquad
\times (2\pi)^3  \delta^{(3)}(\vek{k}+\vek{k}'-\vek{k}_{T})\frac{6 \,f_{\rm NL}}{5}
\nonumber\\
&\qquad\qquad\times 
\left[P_{\mathcal{R}}(k)P_{\mathcal{R}}(k')
+P_{\mathcal{R}}(k')P_{\mathcal{R}}(k_{T})
+P_{\mathcal{R}}(k)P_{\mathcal{R}}(k_{T})
\right]
\nonumber\\
&\qquad\qquad\times 
W_{X}(k,k',z_{\rm rec})
\nonumber\\
&=
\frac{6\,f_{\rm NL}(4\pi)^2}{25}\,(-i)^\ell i^{\ell'} \int\!\! \frac{\id^{3}k_T}{(2\pi)^3}\frac{\id^{3}k}{(2\pi)^3} \,
Y^{*}_{\ell m}(\hat{\vek{k}}_T)\,Y_{\ell' m'}(\hat{\vek{k}}_T)
\nonumber\\
&\qquad\qquad
\times j_{\ell}(k_T r_L)\,j_{\ell'}(k_T r_L)
\nonumber\\
&\qquad\qquad\times 
\left[\{P_{\mathcal{R}}(k)+P_{\mathcal{R}}(k_{T})\}P_{\mathcal{R}}(|\vek{k}_T-\vek{k}|)
+P_{\mathcal{R}}(k)P_{\mathcal{R}}(k_{T})
\right]
\nonumber\\
&\qquad\qquad\times 
W_{X}(k,|\vek{k}_T-\vek{k}|,z_{\rm rec})
\nonumber\\
&=
\frac{12\,f_{\rm NL}(4\pi)^2}{25}\,(-i)^\ell i^{\ell'} \int\!\! \frac{\id^{3}k_T}{(2\pi)^3}\frac{k^2 \id k}{2\pi^2} \,
Y^{*}_{\ell m}(\hat{\vek{k}}_T)\,Y_{\ell' m'}(\hat{\vek{k}}_T)
\nonumber\\
&\qquad\qquad
\times j_{\ell}(k_T r_L)\,j_{\ell'}(k_T r_L)\,P_{\mathcal{R}}(k)P_{\mathcal{R}}(k_{T})\times \mathcal{I}(k, k_T)
\nonumber\\
&=
\frac{12\,f_{\rm NL}(4\pi)^2}{25}\,\frac{\delta_{\ell \ell'}\,\delta_{mm'}}{4\pi} 
\int\!\! \frac{k_T^2 \id k_T}{2\pi^2}\frac{k^2 \id k}{2\pi^2} \,
\nonumber\\
&\qquad\qquad
\times j_{\ell}^2(k_T r_L)\,P_{\mathcal{R}}(k)P_{\mathcal{R}}(k_{T})\times \mathcal{I}(k, k_T),
\end{align}
where we introduced the integral
\beal
\mathcal{I}(k, k_T)
&=\frac{1}{2}\,\int \frac{\id^{2} \hat{k}}{4\pi}
\left[\frac{P_{\mathcal{R}}(|\vek{k}_T-\vek{k}|)}{P_{\mathcal{R}}(k_{T})}
+\frac{P_{\mathcal{R}}(|\vek{k}_T-\vek{k}|)}{P_{\mathcal{R}}(k)}
+1
\right]
\nonumber\\
&\qquad\qquad\times 
W_{X}(k,|\vek{k}_T-\vek{k}|,z_{\rm rec})
\nonumber\\
&=\frac{1}{4}\,\int \id \mu
\left[\frac{P_{\mathcal{R}}(k_1)}{P_{\mathcal{R}}(k_{T})}
+\frac{P_{\mathcal{R}}(k_1)}{P_{\mathcal{R}}(k)}
+1
\right]
W_{X}(k,k_1,z_{\rm rec})
\nonumber\\
&=\frac{1}{2}\,\left[ 
\bar{W}_{X}(k, k,k_T,z_{\rm rec})+\bar{W}_{X}(k_T, k,k_T,z_{\rm rec})\right.
\nonumber\\
&\qquad\qquad\left.+\bar{W}_{X}(k,k_T,z_{\rm rec}) 
\right]
\nonumber\\
\bar{W}_{X}(k,k_T,z_{\rm rec})&=\frac{1}{2}\,\int \id \mu\,W_{X}(k,k_1,z_{\rm rec})
\nonumber\\
&=\frac{1}{2}\,\int_{|k_T-k|}^{k_T+k} \frac{k_1\id k_1}{k k_T}\,W_{X}(k,k_1,z_{\rm rec})
\nonumber\\
\bar{W}_{X}(k_0, k,k_T,z_{\rm rec})
&=\frac{1}{2}\,\int_{|k_T-k|}^{k_T+k} 
\frac{k_1\id k_1}{k k_T}\,\frac{P_{\mathcal{R}}(k_1)}{P_{\mathcal{R}}(k_0)}\,W_{X}(k,k_1,z_{\rm rec})
\end{align}
with $k_1=|\vek{k}_T-\vek{k}|=(k^2_T+k^2-2k_T k \mu)^{1/2}$. Assuming quasi-scale-invariant power spectra one can set  
$\bar{W}_{X}(k_T, k,k_T,z_{\rm rec})\approx (k_T/k)^3\,\bar{W}_{X}(k, k,k_T,z_{\rm rec})\approx 0$ for $k\gg k_T$, making the usual approximation in the ultra-squeezed limit. Even if the small-scale power spectrum amplitude is enhanced to $\Delta^2\simeq 10^{-5}$, which is still allowed by current limits from {\it COBE/FIRAS} \citep{Chluba2012inflaton}, the correction is expected to be tiny.

Since $W_{X}(k,k_1,z_{\rm rec})$ mainly has contributions for $k\simeq k_1$, we can furthermore set $\bar{W}_{X}(k, k,k_T,z_{\rm rec})\approx \bar{W}_{X}(k,k_T,z_{\rm rec})$. This then yields Eq.~\eqref{eq:C_ell_XT} after substitution of $P_{\mathcal{R}}(k)$ and $\bar{W}_{X}(k, k_T,z_{\rm rec})$ for the scale-invariant case. For the scale-dependent case, similar arguments apply and confirm Eq.~\eqref{eq:C_ell_XT}.

\section{Distortion correlation functions}
\label{app:distortion_auto}
To compute the distortion correlation functions we start from
\beal
\label{eq:fourpoint}
\langle (a_{\ell m}^{X})^{*}a_{\ell'm'}^{Y}\rangle
&=
(4\pi)^2\,i^{\ell} (-i)^{\ell'} \!\! \int\!\! 
\frac{\id^{3}k}{(2\pi)^3}\frac{\id^{3}k_1}{(2\pi)^3}\frac{\id^{3}k'}{(2\pi)^{3}}\frac{\id^{3}k_2}{(2\pi)^3} \,
Y_{\ell m}(\hat{\vek{k}}_+)\,Y^{*}_{\ell' m'}(\hat{\vek{k}}'_+)
\nonumber\\
&\qquad\qquad
\times j_{\ell}(k_+ r_L)\,j_{\ell'}(k'_+ r_L)
\nonumber\\
&\qquad\qquad
\times \langle \mathcal{R}(\vek{k})\mathcal{R}(\vek{k}_1) \mathcal{R}(-\vek{k}')\,\mathcal{R}(-\vek{k}_2)\rangle 
\nonumber\\
&\qquad\qquad
\times W_{X}(k,k_1,z_{\rm rec}) W_{Y}(k',k_2,z_{\rm rec})
\end{align}
with $\vek{k}_+=\vek{k}+\vek{k}_1$ and $\vek{k}'_+=\vek{k}'+\vek{k}_2$. We now first evaluate the Gaussian contribution and then consider the non-Gaussian contribution.

\vspace{8mm}
\subsection{Gaussian contribution}
\label{app:distortion_auto_Gaussian}
For the Gaussian contribution to the distortion correlation functions we need
\begin{align}
\label{eq:corr_R}
&\langle \mathcal{R}(\vek{k})\mathcal{R}(\vek{k}_1) \mathcal{R}(-\vek{k}')\,\mathcal{R}(-\vek{k}_2)\rangle
    \nonumber\\
        &\qquad=(2\pi)^6 \delta^{(3)}(\vek{k}_+) \,\delta^{(3)}(\vek{k}'_+) \,P(k)\,P(k')
    \nonumber\\
        &\quad\qquad+(2\pi)^6 \delta^{(3)}(\vek{k}-\vek{k}') \,\delta^{(3)}(\vek{k}_1-\vek{k}_2) \,P(k)\,P(k_1)
    \nonumber\\
        &\qquad\qquad+(2\pi)^6 \delta^{(3)}(\vek{k}-\vek{k}_2) \,\delta^{(3)}(\vek{k}_1-\vek{k}')\,P(k)\,P(k_1).
\end{align}
The first term is associated with the average contribution, $a^{XY}_{00}\neq 0$, for which we find
\beal
\langle (a_{\ell m}^{X})^{*}a_{\ell'm'}^{Y}\rangle^{I}=
4\pi\, & \delta_{\ell\ell'}\delta_{mm'} \delta_{\ell 0} \, \int\!\! \frac{\id^{3}k_1}{(2\pi)^3} \frac{\id^{3}k_2}{(2\pi)^3} 
\,P(k_1)\,P(k_2)
\nonumber\\
&
\qquad \qquad \times 
W_{X}(k_1,k_1,z_{\rm rec})\,
W_{Y}(k_2,k_2,z_{\rm rec})
\\ \nonumber
=
4\pi\, & \delta_{\ell\ell'}\delta_{mm'} \delta_{\ell 0} \,\langle X\rangle\,\langle Y\rangle,
\end{align}
after carrying out the integrals over $\id^3 k$ and $\id^3 k'$. Similarly, by performing the integrals over $\id^3 k'$ and $\id^3 k_2$, for the second contribution we obtain
\beal
\langle (a_{\ell m}^{X})^{*}a_{\ell'm'}^{Y}\rangle^{II}
&=
(4\pi)^2\,i^{\ell} (-i)^{\ell'} \!\! \int\!\! 
\frac{\id^{3}k}{(2\pi)^3}\frac{\id^{3}k_1}{(2\pi)^3} \,
Y_{\ell m}(\hat{\vek{k}}_+)\,Y^{*}_{\ell' m'}(\hat{\vek{k}}_+)
\nonumber\\
&\qquad\qquad
\times j_{\ell}(k_+ r_L)\,j_{\ell'}(k_+ r_L)
\nonumber\\
&\qquad\qquad
\times \,P(k)\,P(k_1)\, W_{X}(k,k_1,z_{\rm rec}) W_{Y}(k,k_1,z_{\rm rec})
\nonumber\\
&=
(4\pi)^2\, i^{\ell} (-i)^{\ell'}\!\! \int\!\! 
\frac{\id^{3}k_+}{(2\pi)^3}\frac{\id^{3}k_1}{(2\pi)^3} \,
Y_{\ell m}(\hat{\vek{k}}_+)\,Y^{*}_{\ell' m'}(\hat{\vek{k}}_+)
\nonumber\\
&\qquad\qquad
\times j_{\ell}(k_+ r_L)\,j_{\ell'}(k_+ r_L)
\nonumber\\
&\qquad\qquad
\times \,P(|\vek{k}_+-\vek{k}_1|)\,P(k_1)
\nonumber\\
&\qquad\qquad
\times\, W_{X}(|\vek{k}_+-\vek{k}_1|,k_1,z_{\rm rec}) W_{Y}(|\vek{k}_+-\vek{k}_1|,k_1,z_{\rm rec})
\nonumber\\
&=
(4\pi)^2\,i^{\ell} (-i)^{\ell'} \!\! \int\!\! 
\frac{\id^{3}k_+}{(2\pi)^3}\frac{k_1^2 \id k_1}{2\pi^2} \,
Y_{\ell m}(\hat{\vek{k}}_+)\,Y^{*}_{\ell' m'}(\hat{\vek{k}}_+)
\\  \nonumber
&\qquad\qquad
\times j_\ell(k_+ r_L)\,j_{\ell'}(k_+ r_L)\,P^2(k_1)\,\bar{W}_{X Y}(k_+, k_1, z_{\rm rec})
\\   \nonumber
&=
4\pi  \delta_{\ell\ell'}\delta_{mm'} \!\! \int\!\! 
\frac{\id k}{k}\!\frac{\id k_1}{k_1} 
 j_{\ell}^2(k r_L)\left[\frac{k^3}{k^3_1}\Delta^4(k_1)\right] \bar{W}_{X Y}(k, k_1, z_{\rm rec})
\end{align}
with azimuthally-averaged power spectrum-weighted window function
\beal
\bar{W}_{X Y}(k, k_1,z_{\rm rec})
&=
\int\!\! \frac{\id^{2}\hat{k}_1}{4\pi}\frac{k_1^3}{|\vek{k}-\vek{k}_1|^3}
\frac{\Delta^2(|\vek{k}-\vek{k}_1|)}{\Delta^2(k_1)}
W_{XY}(k_1,|\vek{k}-\vek{k}_1|,z_{\rm rec})
\nonumber\\
&=
\int\!\! \frac{\id \mu_1}{2}\frac{k_1^3}{k_2^3}
\frac{\Delta^2(k_2)}{\Delta^2(k_1)}
W_{XY}(k, k_1,k_2,z_{\rm rec})
\nonumber\\
&=
\int_{|k-k_1|}^{k+k_1}\!\! \frac{k_2\!\id k_2}{2k k_1}
\frac{k_1^3\Delta^2(k_2)}{k^3_2\Delta^2(k_1)}
W_{XY}(k_1,k_2,z_{\rm rec})
\nonumber\\
&=
\int_{|k-k_1|}^{k+k_1}\!\! \frac{\id k_2}{2 k}
\frac{k_1^2\Delta^2(k_2)}{k^2_2\Delta^2(k_1)}
W_{XY}(k_1,k_2,z_{\rm rec}),
\end{align}
for $W_{XY}(k,k',z)=W_{X}(k,k',z)\,W_{Y}(k,k',z)$ and $k_2=(k^2+k_1^2-2kk_1\mu_1)^{1/2}$.

For the third contribution to the correlation function we find $\langle (a_{\ell m}^{X})^{*}a_{\ell'm'}^{Y}\rangle^{III}=\langle (a_{\ell m}^{X})^{*}a_{\ell'm'}^{Y}\rangle^{II}$. Collecting all terms and defining $\langle (a_{\ell m}^{X})^{*}a_{\ell'm'}^{Y}\rangle=\delta_{\ell\ell'} \delta_{mm'}\,C_\ell^{XY}$ we thus obtain
\beal
\label{eq:Corr_XY_final}
C_\ell^{XY, {\rm G}}&=
4\pi\, \delta_{\ell 0} \,\langle X\rangle\,\langle Y\rangle
\\ \nonumber
&\;+8\pi \!\int\!\! k^2\!\id k\,
 j^2_\ell(k r_L) \int\!\! \frac{\id k_1}{k^4_1}\,\Delta^4(k_1)\,\bar{W}_{X Y}(k, k_1,z_{\rm rec})
\end{align}
for the Gaussian contribution. For a scale-invariant power spectrum, the integral is found to be $C_\ell^{XY, {\rm G}}\simeq 8\pi\,\times 10^{-28}$ for $\ell>0$. This is in good agreement with the result of \citet{Pajer2012} and can usually be neglected. Second-order contributions in $f_{\rm NL}$ to the trispectrum are only larger than the Gaussian contribution for $f_{\rm NL}\gtrsim 1/\sqrt{\Delta(k_T)}\simeq \pot{2}{4}$.

\subsection{Non-Gaussian contribution}
\label{app:distortion_auto_nonGaussian}
To evaluate the non-Gaussian contribution to the distortion correlation function, we use Eq.~\eqref{eq:R4_nonG} in Eq.~\eqref{eq:fourpoint}. The Dirac-$\delta$ function ensures $\vek{k}_+=\vek{k}'_+$, such that after transforming $\id^3 k\rightarrow \id^3 k_+$, transforming $\id^3 k'\rightarrow \id^3 k'_+$ and carrying out the integral over $\id^3 k'_+$, we obtain
\beal
\label{eq:hh2}
\langle (a_{\ell m}^{X})^{*}a_{\ell'm'}^{Y}\rangle
&=
\frac{36}{25}\,(4\pi)^2\,i^{\ell} (-i)^{\ell'} \!\! \int\!\! 
\frac{\id^{3}k_+}{(2\pi)^3}\frac{\id^{3}k_1}{(2\pi)^3}\frac{\id^{3}k_2}{(2\pi)^3} \,
Y_{\ell m}(\hat{\vek{k}}_+)\,Y^{*}_{\ell' m'}(\hat{\vek{k}}_+)
\nonumber\\
&\quad
\times j_{\ell}(k_+ r_L)\,j_{\ell'}(k_+ r_L)
\nonumber\\
&\quad
\times \mathcal{P}(k_+,k_1,k_2,|\vek{k}_1-\vek{k}_+|,|\vek{k}_2-\vek{k}_+|,|\vek{k}_1+\vek{k}_2-\vek{k}_+|,|\vek{k}_1-\vek{k}_2|) 
\nonumber\\
&\quad
\times W_{X}(|\vek{k}_1-\vek{k}_+|,k_1,z_{\rm rec}) W_{Y}(|\vek{k}_2-\vek{k}_+|,k_2,z_{\rm rec}).
\end{align}
Explicitly, $\mathcal{P}$ reads
\beal
\mathcal{P}&=P(|\vek{k}_1-\vek{k}_+|)\,P(k_1)
[P(|\vek{k}_2-\vek{k}_1|)+P(|\vek{k}_1+\vek{k}_2-\vek{k}_+|)] f_{\rm NL}(|\vek{k}_2-\vek{k}_+|) f_{\rm NL}(k_2)
\nonumber\\
&
+P(|\vek{k}_1-\vek{k}_+|)\,P(|\vek{k}_2-\vek{k}_+|)
[P(k_+)+P(|\vek{k}_1+\vek{k}_2-\vek{k}_+|)] f_{\rm NL}(k_1) f_{\rm NL}(k_2)
\nonumber\\
&
+P(|\vek{k}_1-\vek{k}_+|)\,P(k_2)[P(|\vek{k}_2-\vek{k}_1|)+P(k_+)] f_{\rm NL}(|\vek{k}_2-\vek{k}_+|) f_{\rm NL}(k_1)
\nonumber\\
&
+P(|\vek{k}_2-\vek{k}_+|)\,P(k_1)[P(|\vek{k}_2-\vek{k}_1|)+P(k_+)] f_{\rm NL}(|\vek{k}_1-\vek{k}_+|) f_{\rm NL}(k_2)
\\ \nonumber
&
+P(|\vek{k}_2-\vek{k}_+|)\,P(k_2)
[P(|\vek{k}_2-\vek{k}_1|)+P(|\vek{k}_1+\vek{k}_2-\vek{k}_+|)] f_{\rm NL}(|\vek{k}_1-\vek{k}_+|) f_{\rm NL}(k_1)
\\ \nonumber
&
+P(k_1)\,P(k_2)
[P(k_+)+P(|\vek{k}_1+\vek{k}_2-\vek{k}_+|)] f_{\rm NL}(|\vek{k}_1-\vek{k}_+|) f_{\rm NL}(|\vek{k}_2-\vek{k}_+|).
\end{align}
The angle-averages over $\id^2 \hat{k}_1$ and $\id^2 \hat{k}_2$ in Eq.~\eqref{eq:hh2} make the integrant independent of $\id^2 \hat{k}_+$ aside from in the arguments of the spherical harmonics. We can thus write
\beal
\label{eq:hh3}
\langle (a_{\ell m}^{X})^{*}a_{\ell'm'}^{Y}\rangle
&=
4\pi \,\frac{36}{25}\,\delta_{\ell\ell'}\delta_{mm'} \!\! \int\!\! 
\frac{k_+^2\id k_+}{2\pi^2}\times j_{\ell}^2(k_+ r_L) \frac{\id^{3}k_1}{(2\pi)^3}\frac{\id^{3}k_2}{(2\pi)^3} 
\nonumber\\
&\quad
\times \mathcal{P}(k_+,k_1,k_2,|\vek{k}_1-\vek{k}_+|,|\vek{k}_2-\vek{k}_+|,|\vek{k}_1+\vek{k}_2-\vek{k}_+|,|\vek{k}_2-\vek{k}_1|) 
\nonumber\\
&\quad
\times W_{X}(|\vek{k}_1-\vek{k}_+|,k_1,z_{\rm rec}) W_{Y}(|\vek{k}_2-\vek{k}_+|,k_2,z_{\rm rec}),
\end{align}
where $\vek{k}_+$ now defines the $z$-axis. Due to the Bessel function, $k_+$ is restricted to small values (long-wavelength mode). Similarly, the distortion window functions require $k_+\ll k_1$ and $k_+\ll k_2$. We have seen that this approximation works extremely well for the previous cases, so that we find 
\bsub
\label{eq:hh4}
\beal
\langle (a_{\ell m}^{X})^{*}a_{\ell'm'}^{Y}\rangle
&\approx
4\pi \,\frac{36}{25}\,\delta_{\ell\ell'}\delta_{mm'} \!\! \int\!\! 
\frac{k_+^2\id k_+}{2\pi^2}\times j_{\ell}^2(k_+ r_L) \frac{\id^{3}k_1}{(2\pi)^3}\frac{\id^{3}k_2}{(2\pi)^3} 
\nonumber\\
&\quad
\times \mathcal{P}(k_+,k_1,k_2,|\vek{k}_1+\vek{k}_2|,|\vek{k}_2-\vek{k}_1|) 
\nonumber\\
&\quad
\times W_{X}(k_1,k_1,z_{\rm rec}) W_{Y}(k_2,k_2,z_{\rm rec})
\\[2mm] \nonumber
\mathcal{P}&\approx 
4P(k_+)\,P(k_1)\,P(k_2) \, f_{\rm NL}(k_1) f_{\rm NL}(k_2)
\nonumber\\
&
\quad+2P(k_1)\,P(k_2) [P(|\vek{k}_1+\vek{k}_2|)+P(|\vek{k}_2-\vek{k}_1|)] f_{\rm NL}(k_1) f_{\rm NL}(k_2)
\nonumber\\
&
\qquad+2\,P^2(k_1)[P(|\vek{k}_2-\vek{k}_1|)+P(|\vek{k}_1+\vek{k}_2|)] f_{\rm NL}^2(k_2).
\end{align}
\esub
In the expression for $\mathcal{P}$, those terms $\propto P(k_+)$ dominate, such that
\beal
\label{eq:hhf}
\langle (a_{\ell m}^{X})^{*}a_{\ell'm'}^{Y}\rangle
&\approx
4\pi \,\frac{144}{25}\,\delta_{\ell\ell'}\delta_{mm'} \!\! \int\!\! 
\frac{k_+^2\id k_+}{2\pi^2}\times j_{\ell}^2(k_+ r_L) P(k_+)\,
\nonumber\\
&\quad
\times \int \frac{k_1^2\id k_1}{2\pi^2}\frac{k_2^2\id k_2}{2\pi^2} 
 P(k_1)\,P(k_2) \, f_{\rm NL}(k_1) f_{\rm NL}(k_2) 
\nonumber\\
&\qquad\qquad
\times W_{X}(k_1,k_1,z_{\rm rec}) W_{Y}(k_2,k_2,z_{\rm rec}).
\end{align}
As this expression already indicates, for large values of $f_{\rm NL}$, this contribution can be significantly larger than the Gaussian contribution \citep{Pajer2012}.

\end{appendix}

\end{document}